\documentclass[aoas,preprint]{imsart}

\RequirePackage[OT1]{fontenc}
\RequirePackage{amsthm}
\RequirePackage[intlimits]{amsmath}
\RequirePackage{natbib}
\RequirePackage[colorlinks,citecolor=blue,urlcolor=blue]{hyperref}
\usepackage[utf8]{inputenc}
\usepackage{amssymb, amsfonts}
\usepackage{graphicx,psfrag,epsf}
\usepackage{graphics}
\usepackage{enumerate}
\usepackage{booktabs}
\usepackage{multirow}
\usepackage{color}
\usepackage{bm}
\usepackage{bbm}
\usepackage{parskip}
\usepackage{tikz}
\usepackage{geometry}
\usepackage[export]{adjustbox}
\usepackage{url} 
\usepackage{caption}
\usepackage{subcaption}
\usepackage{mathtools}
\usepackage{floatrow}
\usepackage[toc,page]{appendix}
\usepackage{verbatim}
\usepackage{pdfpages}
\usepackage{lscape}
\usepackage{lineno}
\usepackage{array}

\arxiv{arXiv:0000.0000}

\startlocaldefs

\newcommand{\bbeta}{ \mbox{\boldmath $\beta$}}

\newcommand{\bdelta}{ \mbox{\boldmath $\delta$}}

\newcommand{\bomega}{ \mbox{\boldmath $\omega$}}

\newcommand{\ba}{ \mbox{\bf a}}

\newcommand{\bs}{ \mbox{\bf s}}
\newcommand{\bb}{ \mbox{\bf b}}

\newcommand{\iid}{\stackrel{iid}{\sim}}

\newcommand{\PM}{ \mbox{PM$_{2.5}$}}

\newcommand{\bbi}{\mbox{$\mathbb{I}$}}

\newcommand{\beq}{ \begin{equation}}
\newcommand{\eeq}{ \end{equation}}
\newcommand{\beqn}{ \begin{eqnarray}}
\newcommand{\eeqn}{ \end{eqnarray}}

\newcommand{\rtwo}{\mbox{${\mathbb{R}^{2}}$}}

\newcommand{\rpm}{\sbox0{$1$}\sbox2{$\scriptstyle\pm$}
  \raise\dimexpr(\ht0-\ht2)/2\relax\box2 }

\def\Lp{\left(}
\def\Rp{\right)}

\numberwithin{equation}{section}
\theoremstyle{plain}

\endlocaldefs

\begin{document}

\begin{frontmatter}
\title{Statistical downscaling with spatial misalignment: Application to wildland fire $\PM$ concentration forecasting}
\runtitle{Statistical downscaling with spatial misalignment}
\begin{aug}
\author{\fnms{Suman} \snm{Majumder}\corref{c1}\thanksref{m1}\ead[label=e1]{smajumd2@ncsu.edu}},
\author{\fnms{Yawen} \snm{Guan}\thanksref{m2}\ead[label=e2]{yguan12@unl.edu}
\ead[label=u1]{https://sites.google.com/a/ncsu.edu/yguan/}},
\author{\fnms{Brian~J.} \snm{Reich}\thanksref{m1}\ead[label=e3]{bjreich@ncsu.edu}\ead[label=u2]{https://www4.stat.ncsu.edu/~reich/}}
\author{\fnms{Susan} \snm{O'Neill}\thanksref{m3}}
\and
\author{\fnms{Ana~G.} \snm{Rappold}\thanksref{m4}\ead[label=e4]{rappold.ana@epa.gov}}

\runauthor{Majumder et al.}

\affiliation{North Carolina State University\thanksmark{m1}, University of Nebraska\thanksmark{m2}, United States Forest Service, Pacific Northwest Research Station\thanksmark{m3} and United States Environmental Protection Agency\thanksmark{m4}}

\address{Suman~Majumder\\
North Carolina State University\\
Department of Statistics\\
2311 Stinson Drive\\
Raleigh, NC 27695\\
\printead{e1}}

\address{Yawen~Guan\\
University of Nebraska\\
Department of Statistics\\
340 Hardin Hall North Wing\\
Lincoln, NE 68583\\
\printead{e2}\\
\printead{u1}}

\address{Brian~J.~Reich\\
North Carolina State University\\
Department of Statistics\\
Campus Box 8203\\
5212 SAS Hall\\
2311 Stinson Drive\\
Raleigh, NC 27695\\
Phone: (919) 513 - 7686\\
\printead{e3}\\
\printead{u2}}

\address{Susan~O'Neill\\
400 N 34th Street, Suite 201\\
Seattle, WA 98103\\
}

\address{Ana~G.~Rappold\\
101 Manning Drive\\
Chapel Hill, NC 27514\\
\printead{e4}}
\end{aug}\footnote[5]{The views expressed in this manuscript are those of the individual authors and do not necessarily reflect the views and policies of the U.S. Environmental Protection Agency. Mention of trade names or commercial products does not constitute endorsement or recommendation for use.}

\begin{abstract}
Fine particulate matter, PM$_{2.5}$, has been documented to have adverse health effects and wildland fires are a major contributor to $\PM$ air pollution in the US. Forecasters use numerical models to predict PM$_{2.5}$ concentrations to warn the public of impending health risk. Statistical methods are needed to calibrate the numerical model forecast using monitor data to reduce bias and quantify uncertainty. Typical model calibration techniques do not allow for errors due to misalignment of geographic locations. We propose a spatiotemporal downscaling methodology that uses image registration techniques to identify the spatial misalignment and accounts for and corrects the bias produced by such warping. Our model is fitted in a Bayesian framework to provide uncertainty quantification of the misalignment and other sources of error. We apply this method to different simulated data sets and show enhanced performance of the method in the presence of spatial misalignment. Finally, we apply the method to a large fire in Washington state and show that the proposed method provides more realistic uncertainty quantification than standard methods.
\end{abstract}



\end{frontmatter}


\section{Introduction}\label{s:intro}

Air pollution associated with wildland fire smoke is an increasingly pressing health concern \citep{Dennekamp2011,Rappold2011,Johnston2012,Dennekamp2015,Haikerwal2015,Haikerwal2016,Wettstein2018}.  Reliable short-term forecasts of fire-associated health risk using numerical models facilitate informed decision making for local populations. Numerical models produce forecasts on a course grid and are prone to bias.   Assimilating point-level monitor data with numerical-model output can reduce bias and provide more realistic uncertainty quantification \citep[e.g.,][]{berrocal2010bivariate, berrocal2010spatio, kloog2011assessing,zhou2011calibration, zhou2012estimating, berrocal2012space, reich2014spectral,chang2014calibrating}. However, most downscaling methods only correct for additive and scaling biases and fail to guard against spatial misalignment errors. This is problematic for wildland fire smoke forecasting because a common source of error is in predicting the direction of the fire plume which cannot be accounted for by additive and scaling correction to the forecast. This motivates us to develop a statistical downscaling method that accounts for spatial misalignment errors.

Spatial misalignment correction can be achieved using standard  image registration (or warping) techniques, ranging from simple affine and polynomial transformations to more sophisticated methods such as Fourier based transforms \citep{kuglin1975phase, de1987registration}, nonparametric approaches like elastic deformation \citep{burr1981dynamic,tang1993image, barron1994performance} and thin-plate splines \citep{bookstein1989principal,mardia1994image,mardia1996kriging}. Besides image processing and medical imaging, warping is also popular in speech processing \citep{sakoe1978dynamic}, handwriting analysis \citep{burr1983designing}, determination of alignment of boundaries of ice floes \citep{mcconnell1991psi} and more recently, weather forecast analysis \citep{hoffman1995distortion, alexander1999effect,sampson1999operational,reilly2004using, gilleland2010analyzing}. 

\cite{sampson:1992} used warping of spatial coordinates to model non-stationary and non-isotropic spatial covariance structures.~\cite{anderes2008estimating} and \cite{anderes2009consistent} developed methods for estimating deformation of isotropic Gaussian random fields. The first attempt at using warping for forecast verification in statistics was proposed by \cite{aberg2005image}. Image warping in wind field modelling was proposed by \cite{ailliot2006autoregressive} and \cite{fuentes2008spatial} used warping to assimilate two different sources of rainfall data in a single model. \cite{kleiber2014model} used warping in the context of model emulation and calibration framework. They assume the observations lie on a grid and the spatial features are completely observed so that standard image registration techniques such as landmark registration can be used for estimating the warping function. However, this approach does not apply to our downscaling problem because the monitoring stations are spatially sparse and the shape and direction of the fire plume are not observed. The estimation of the warping function is challenging and further complicated by the dynamic environment, such as changes in the wind pattern.

 We propose a new statistical downscaling method that optimizes the information from available forecasts and real-time monitoring data. We achieve this through (1) introducing a warping function to allow for flexible model discrepancy beyond the additive and multiplicative biases and (2) multi-resolution modeling to use only the appropriate spatial resolution to inform prediction. We estimate the spatial misalignment between the forecast and the observed data using a penalized B-spline approach. We also use spectral smoothing \citep{reich2014spectral} to capture important patterns more vividly and reduce noise simultaneously. By coalescing these two methods in a single model, we propose a novel downscaling model that accounts for spatial misalignment as well as the usual additive and scaling biases while smoothing out the forecast to improve prediction.

The remainder of the paper proceeds as follows.  Section \ref{s: data_des} introduces the motivating dataset and Section \ref{s:model} describes the proposed method.  The performance of the model and its component models are studied extensively using a simulation study in Section \ref{s:sim}.  The method is applied to forecasting air pollution during a major fire in Washington State in Section \ref{s:data}, where we show that accounting for spatial misalignment provides better assessment of uncertainty.  Section \ref{s:disc} concludes.

\section{$\PM$ Data for Washington State} \label{s: data_des}

We have two sources of $\PM$ data: numerical model forecasts on a grid and ground monitoring station scattered around the state. Both data sets give hourly $\PM$ measurements for the state of Washington from August 13, 2015 to September 16, 2015, a period with severe wildland fires.

The numerical forecasts were generated by the BlueSky modeling system on a $4$km $\times$ $4$km grid resulting in a $200 \times 95$ grid covering Washington. The model is run daily at midnight and provides an hourly forecast for the next $84$ hours of which we use only the forecasts for the first $24$ hours for our analysis. The model only forecasts $\PM$ levels created by wildland fires and does not contain any information about $\PM$ generated from other sources such as traffic or industry.

The second source of data is from the ground monitoring stations which measure the total $\PM$ level at the corresponding locations. We have $55$ monitor stations throughout the state of Washington. These monitors include both permanent and temporary monitors that are placed near the areas expected to be impacted by the fire. Approximately $7\%$ of the observations are missing.

\begin{figure}
    \centering
    \includegraphics[height = 0.35\textheight]{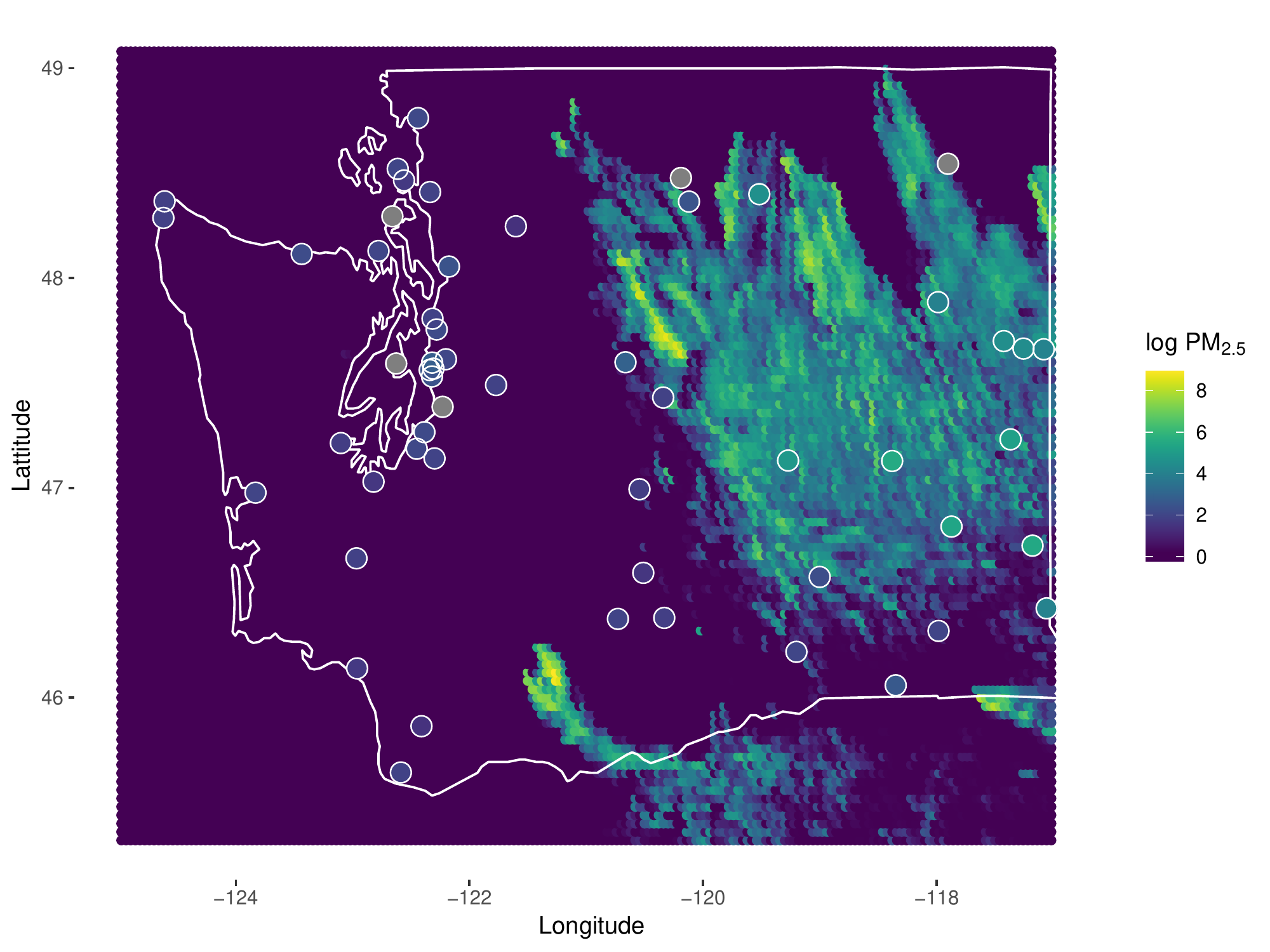}
    \caption{$\PM$ concentration in $\mu g/m^3$ (log scale) in Washington at 04:00 GMT on August 22, 2015. The background map shows the forecast from the numerical model, while the circle shows the location of the monitoring station. The color of the circle  indicates the concentration level of the observed $\PM$ in log scale with missing value colored in gray.}
    \label{fig:data_des}
\end{figure}

Figure \ref{fig:data_des} shows the concentration of $\PM$ $\Lp \log (1+\PM) \Rp$ (in $\mu g/m^3$) on 22 August, 2015 at 04:00 GMT. The circles indicate the locations of the stations and their colors correspond to the observed $\log \PM$ values; missing observations are colored in gray. The background map shows the numerical model forecast. There is obvious difference in the spatial resolution of the two sources of data as well as the type of the data. The numerical model forecast only give information about wildland fire $\PM$ emissions, whereas the monitor station data includes both wildland fire $\PM$ emission and $\PM$ emission from other sources. This adds an additional level of difficulty to model and infer about the same phenomenon from the two sources of data.

\section{Statistical model}\label{s:model}

Let $Y_t(\bs)$ denote the measured $\log \PM$ from the monitor at spatial location $\bs=(s_1,s_2)^{\sf T}$ on day $t$, and $X_t(\bs)$ be corresponding numerical forecast.  Instead of directly relating these variables, we associate $Y_t(\bs)$ to a smoothed and warped forecast to account for model discrepancies.  Let $w:  \rtwo \rightarrow \rtwo$ be a warping function that maps $\bs$ to a new location $w(\bs) = \Lp w_1(\bs), w_2(\bs) \Rp^{\sf T}$ and $\tilde{X}_{t}(\bs)$ to be smoothed forecast.  The model is then 
\beq 
\label{eq : likelihood}
Y_t(\bs) = \beta_0(\bs) + \beta\tilde{X}_{t}(w(\bs))+ \epsilon_t(\bs),
\eeq where  $\epsilon_t(\bs)\iid\mbox{Normal}(0,\sigma^2)$ is error. Since the smoothed and warped forecast is a product of a atmospheric dispersion model that already takes into account the spatiotemporal variability as well as the effects of meteorological components and other factors, we assume that the errors $\epsilon_t(\bs)$ are independent over space and time. 

\subsection{Model for the Spatially Varying Intercept} \label{ss: beta0}

A spatially varying intercept is employed to correct for possible additive bias. In our motivating example, additive bias in the monitor station observations come from other sources of $\PM$, such as traffic and industry that are not included in the numerical forecast. We model the spatially varying intercept using finite basis function expansion \beq 
\label{eq: var_beta} \beta_0(\bs) = b_0 + \sum_{j=1}^J\sum_{k=1}^K A^0_j(s_1)B^0_k(s_2)b_{jk}.\eeq
We use known basis functions for the two coordinates, $A^0_j(s_1)$ and $B^0_k(s_2)$, and estimate the coefficients $b_{jk}$ and $b_0$. Although other choices of basis functions are possible, we use an outer product of B-spline basis functions, that is, $A^0_j(s_1)$ and $B^0_k(s_2)$ are univariate B-spline basis functions with $J$ and $K$ knots, respectively. Cubic B-splines basis functions are a sensible choice as they can approximate any smooth function in a bounded domain.

A natural problem in finite basis function expansion based modelling is the choice of number of knots and their position. We select $J$ and $K$ to be large enough to capture the variability in the data with enough detail and use a penalized B-spline approach to prevent overfitting. Penalization is achieved by employing a Gaussian prior distribution on the coefficients $\bb = (b_{11}, b_{12}, \ldots b_{JK})^{\sf T}$ with mean $\mathbf{0}$ and covariance $\sigma^2_0 \Sigma_0$. $\Sigma_0$ has a conditional autoregressive (CAR) covariance structure, i.e, $\Sigma_0 = \Lp M_0 - \rho_0E_0\Rp^{-1}$, where $E_0$ is the adjacency matrix for the coefficients in $\bb$ and $M_0$ is a diagonal matrix with the number of neighbors for each knot on the diagonal. The coefficients $b_{jk}$ and $b_{j'k'}$ are considered neighbors if $|j-j'| + |k-k'| = 1$. 

\subsection{Model for the Warping Function} \label{ss: warp_model}
 We approximate the warping function using the finite basis function expansion
\beq
\label{eq: warpdef}
w_l(\bs) = s_l + \sum_{j=1}^{J_1}\sum_{k = 1}^{J_2}A_j(s_1)B_k(s_2)a_{jkl} \,,\ l=1,2. \eeq
The warping function is defined by basis functions for the two coordinates $A_j(s_1)$ and $B_{k}(s_2)$  and the corresponding coefficients $a_{jkl}$. We use an outer product of B-spline basis functions for our model here as well, that is, $A_j(s_1)$ and $B_k(s_2)$ are univariate cubic B-spline basis functions with $J_1$ and $J_2$ knots, respectively. However, B-spline would not be a good choice if the warped location is outside the bounded domain. This is tackled by forcing any point outside the grid to be remapped to its closest point on the grid boundary.

Other applications of warping in spatial statistics have used some restrictions on the form of the warping function. For example, \cite{sampson:1992} restricted the class of warping functions to one-to-one functions and \cite{snelson2004warped} restricted the warping functions to be monotone and have the entire real line as its range. Such restrictions are not necessary here since warping the space for covariates does not present problems of preserving measure-theoretic properties or positive definiteness of the covariance structure. Therefore, we can apply warping functions that might map multiple locations to one point in the warped image. This may be unavoidable if the forecast is available only on a course spatial grid and multiple monitors reside in the same grid cell.

While insisting that the warping function is one-to-one is unnecessary and overly restrictive, we do impose a prior penalty to avoid overfitting. Our prior encourages the warping function to be smooth and centered around identity warp, $w(\bs) = \bs$. We consider identical priors for the coefficients $\ba_l = \Lp a_{11l}, \ldots, a_{J_1J_2l} \Rp$ for each $l=1,2$ and that $\ba_1$ and $\ba_2$ are independent. To ensure a smooth warping function, we use a spatial prior for $\ba_l$ defined as a  neighboring scheme based on the indices that involves the rook neighbors for each index when viewed to be placed on a two-dimensional integer grid. That is, $a_{jkl}$ and $a_{j'k'l}$ are neighbors if $|j-j'| + |k - k'| = 1$. A correlation structure for such a neighboring scheme is created by assigning a CAR covariance structure $\mathbf{\Sigma}_w = (\mathbf{M}_1-\rho_w\mathbf{E}_1)^{-1}$ to the normally distributed coefficients, with $\mathbf{E}_1$ being the adjacency matrix and $\mathbf{M_1}$ being the diagonal matrix with $i$th diagonal entry equal to the number of neighbors of the $i$th point. This means that $\ba_l$ has a Gaussian distribution with mean $\mathbf{0}$ and covariance $\sigma_a^2\mathbf{\Sigma}_w$. By setting $E(\ba) = \mathbf{0}$, we shrink the warping function towards the identity function.

\subsection{Model for the Smoothing Function} \label{ss: smoothing_model}
Smoothing the forecast eliminates spurious small-scale variation and allows align large-scale features of the forecast such as smoke plumes with the monitor data. Since the forecast is on a regular grid, the smoothing can be achieved using the spectral downscalar proposed by \cite{reich2014spectral}. The spectral representation of the forecast is \beq \label{eq: xfou}
X_t(\bs) = \int \exp \Lp -i\bomega^{\sf T}\bs \Rp Z_t(\bomega) d\bomega, \eeq where $\bomega \in \rtwo$ is a frequency and \beq \label{eq : xinvfou}
Z_t(\bomega) = \int \exp (i\bomega^{\sf T}\bs)X_t(\bs) d\bs\eeq
is the inverse Fourier transform of the forecast.  This decomposes the forecast's signals at different frequencies $Z_t(\bomega)$. Processes that comprises of lower frequencies contain the information about the large-scale patterns, while processes corresponding to higher frequencies holds local information. We capture the forecast features at different resolutions using \beq \label{eq : xlayer}
\tilde{X}_{lt}(\bs) = \int V_{l}(\bomega) \exp (-i\bomega^{\sf T}\bs)Z_{t}(\bomega) d\bomega,\eeq where $V_{l}(\bomega)$ are known basis functions that serve as weights based on frequencies satisfying $\int V_{l}(\bomega) d\bomega = 1 \,,\ \forall l$. A useful choice for the basis functions are Bernstein polynomials, as suggested by \cite{reich2014spectral} (see the \href{subsec: smooth basis}{Appendix} for details). We then reconstruct the smoothed process by \beq \tilde{X}_t(\bs) = \sum_{l=1}^L \alpha_l\tilde{X}_{lt}(\bs).\eeq Smoothing is achieved if $\alpha_l\approx 0$ for terms with large $||\bomega||$, and if $\alpha_l = 1$ for all $l$, the smoothed forecast reduces to the original forecast, i.e, $\tilde{X}_t(\bs) = X_t(\bs)$

Constructing $\tilde{X}_{lt}(\bs)$ requires computing the stochastic integrals in (\ref{eq : xinvfou}) and (\ref{eq : xlayer}). For fast computing, these integrals are approximated using two dimensional discrete Fourier transform and inverse discrete Fourier transform as \beq
    Z_{pt} \approx \frac{1}{P}\sum_{q = 1}^P  \exp(i\bomega_p^{\sf T}\bs_q) X_t(\bs_q) \,\ \text{and} \,\ \tilde{X}_{lt}(\bs) \approx \sum_{p = 1}^P V_{l}(\bomega_p) \exp (-i\bomega_p^{\sf T}\bs)Z_{pt},
\eeq where the forecast is on a grid of $P_1 \times P_2$ and $P = P_1P_2$. 

In \ref{eq : likelihood}, the scale of $\beta$ and $\alpha_1, \alpha_2, \ldots, \alpha_L$ are not identified, so we reparametrize to $\bbeta = \beta (\alpha_1, \alpha_2, \ldots, \alpha_L)^{\sf T} = \Lp \beta_1, \beta_2, \ldots, \beta_L \Rp^{\sf T}$ and place a prior on $\bbeta$. To prevent overfitting, we use the same penalized splines approach as before. We put another CAR covariance structure on $\bbeta$ with the neighboring scheme based on their indices, as before, \[\bbeta \sim N\Lp \mathbf{0}, \sigma^2 \tau^2\mathbf{D}_x \Rp ,\]where $\mathbf{0}$ is the zero vector of length $L$ and $\mathbf{D}_x = (\mathbf{M}_2 - \rho_x\mathbf{E}_2)^{-1}$ is the corresponding CAR covariance structure and $\mathbf{E}_2$ being the adjacency matrix with terms $l$ and $k$, considered neighbors if $|l-k|=1$ and $\mathbf{M}_2$ being the corresponding diagonal matrix created similarly as before. 

\subsection{Model Details} \label{subsec: m_d}

Since the forecast, and the spectral covariates, can only be computed on a grid and the monitoring sites are non-gridded points in $\rtwo$, we use the nearest grid neighbor as a proxy for forecast at the station. That is, we use the model
\[Y_t(\bs) = \beta_0(\bs) + \sum_{l=1}^L\beta_l\tilde{X}_{lt}(\tilde{w}(\bs)) + \epsilon_t(\bs),\]
where $\tilde{w}(\bs)$ is the location of the closest forecast grid cell to $w(\bs)$. Any point that goes outside the grid as a result of the warping is set at the nearest grid point, as mentioned earlier.
 
To complete the Bayesian model we specify the priors for the hyperparameters: $\sigma^2_0 \sim IG(0.01,0.01)$ and $\sigma^2 \sim IG(0.01,0.01)$. We assume $\sigma_a^2 \sim$Half-Normal$(0.15)$. This sets the $99$th percentile for the prior to be 1. This choice of prior strongly suggest the warping to be adequately smooth. We put a Beta(10,1) prior on the hyperparameters $\rho_0, \rho_a$ and $\rho_x$, suggesting a minimal level of spatial correlation being present. Instead of choosing a hyperprior for $\tau^2$, we set $\tau^2 = 10$. This helps avoid numerical instability in the computational process and yet provides enough prior uncertainty for the $\bbeta$ parameter. We recommend choosing $J, K, J_1$ and $J_2$ to be large, e.g., so that the number of basis functions is roughly the same as the number of monitor stations, as the penalization should set the unnecessary coefficients to zero, thus reducing it to a simpler model. We use $(J,K) = (J_1, J_2) = (6,4), (10,5)$ or $(12,8)$ for our simulation study scenarios with the corresponding number of monitor stations being $25, 50$ or $100$. For the data example, we use $(J,K) = (J_1,J_2) = (11,5)$. We should chose $L$ to be large as well since we added a penalization for that too. However, in this case we do not need to choose $L$ to be as large as the number of monitor stations. For instance, we use $L = 15$ throughout our studies and data example.


\section{Simulation Study}\label{s:sim}

In this section, we conduct a simulation study to explore the performance of the proposed method in different scenarios. We consider four data generation processes and three sets of monitor station locations for each of these processes and create $30$ datasets for each combination of these factors. We use the forecast from the dataset described in Section \ref{s: data_des} for August 18, 2015 to August 22, 2015 as $X_t(\bs)$. The grid size for the simulation study was therefore the same as the forecast grid of the data, $200 \times 95$.

\begin{figure}
\begin{subfigure}{.49\textwidth}
    \flushleft
    \includegraphics[width = \textwidth]{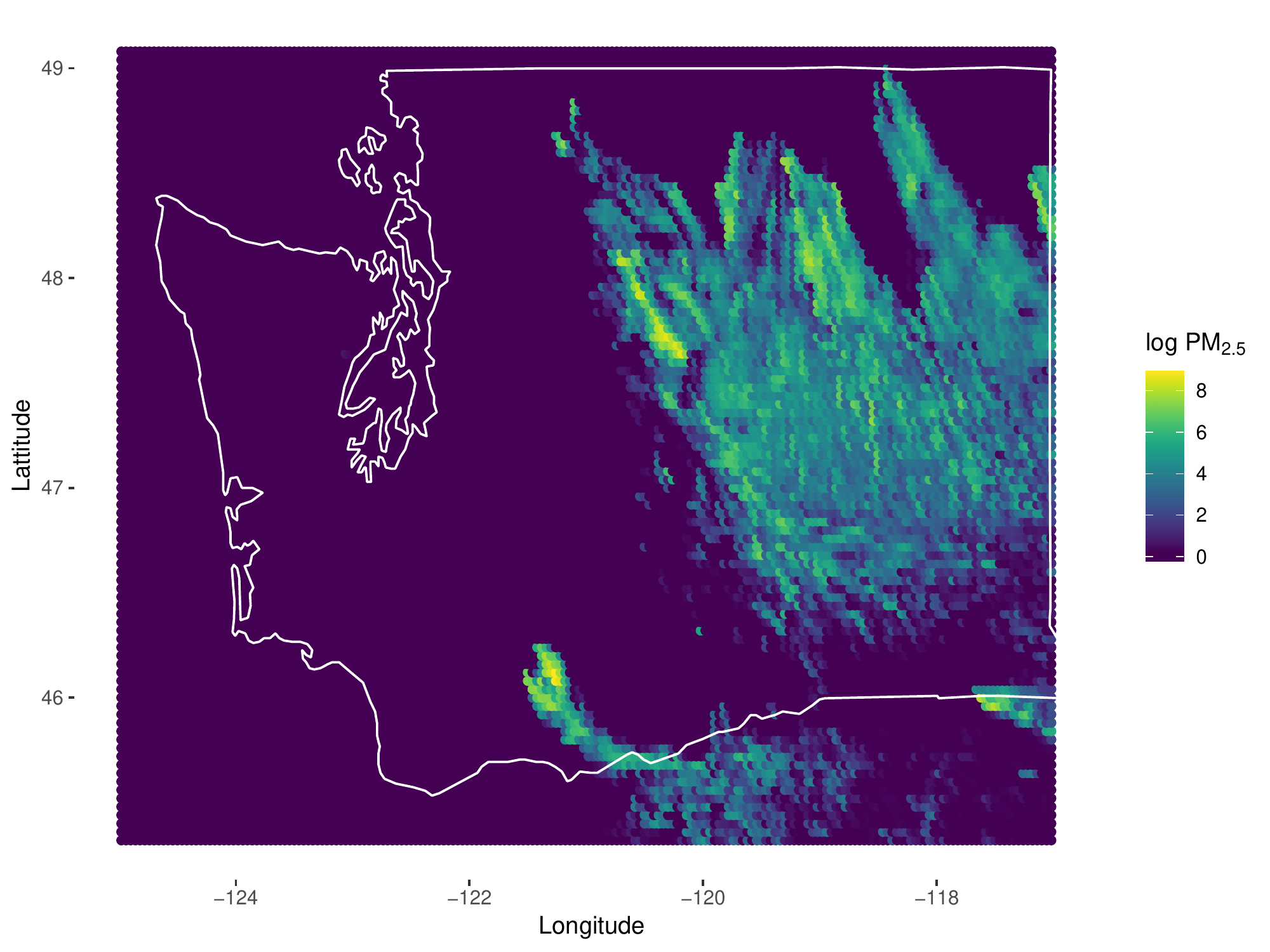}
\end{subfigure}
\begin{subfigure}{.49\textwidth}
    \flushright
    \includegraphics[width = \textwidth]{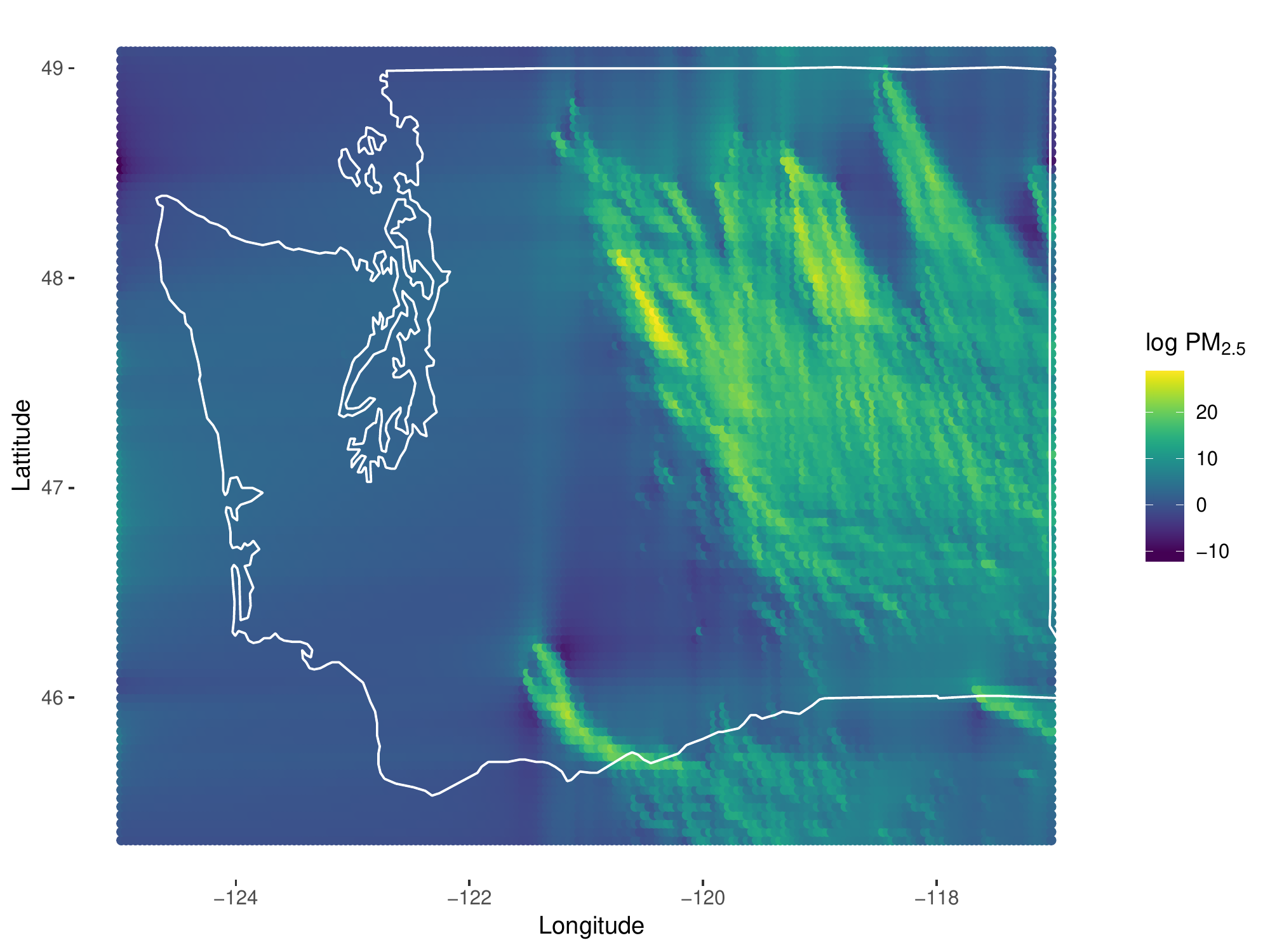}
\end{subfigure}
\begin{subfigure}{.49\textwidth}
    \flushleft
    \includegraphics[width = \textwidth]{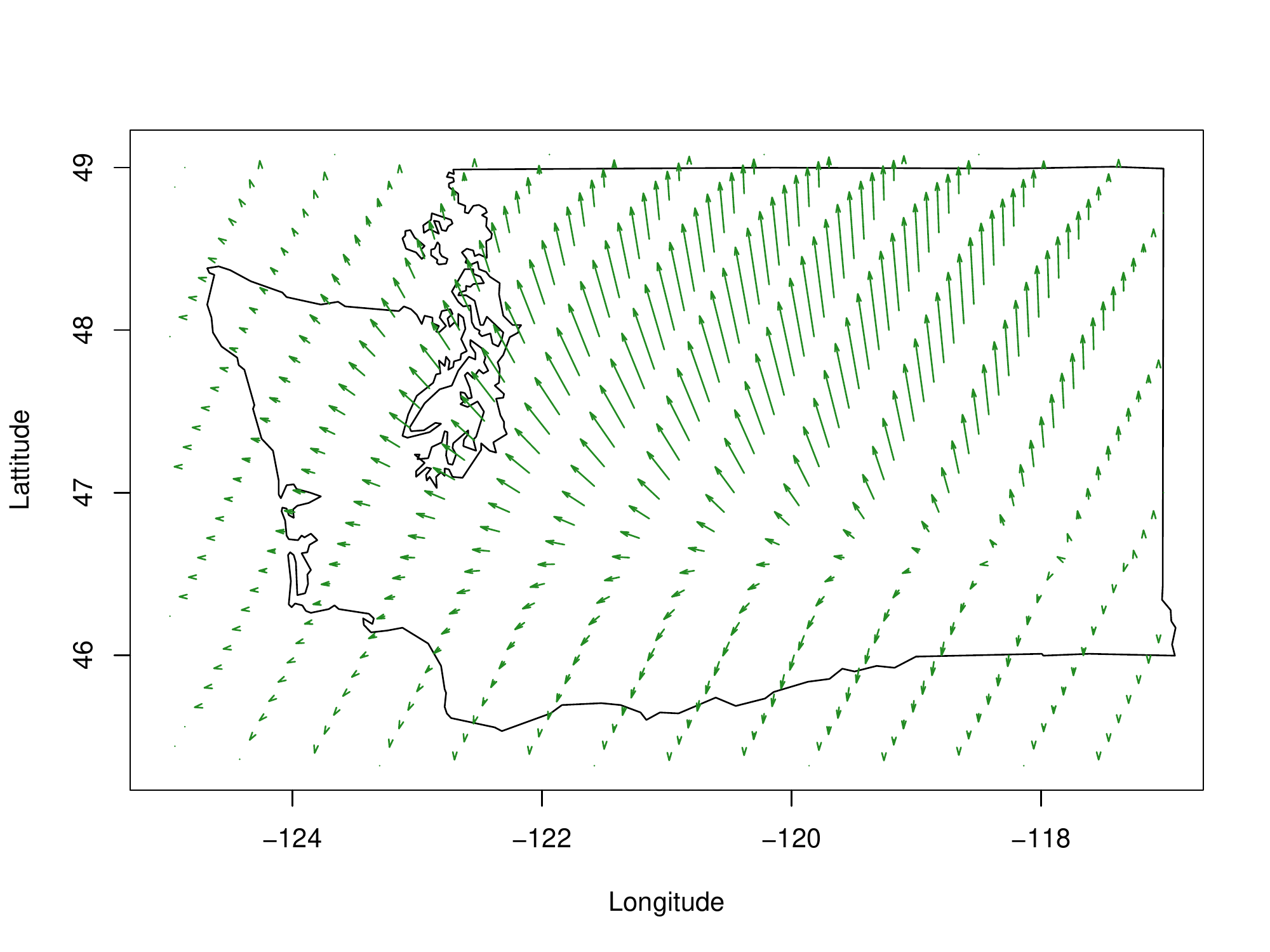}
\end{subfigure}
\begin{subfigure}{.49\textwidth}
    \flushright
    \includegraphics[width=\textwidth]{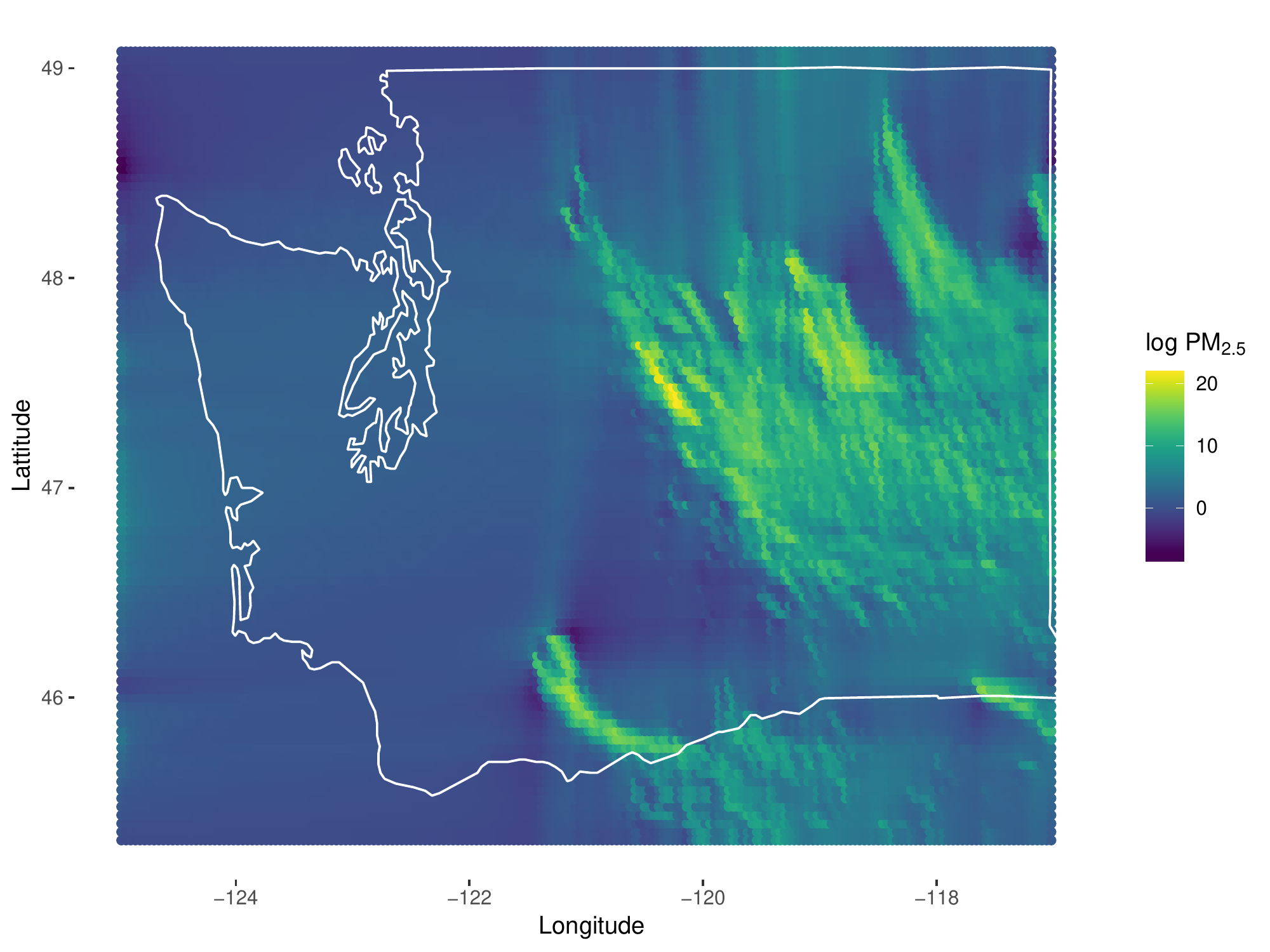}
\end{subfigure}
\caption{Original forecast $X_t(\bs)$ in $\log$-scale on August $22$, 2015 at $4:00$ AM (top left); The smoothed forecast $\tilde{X}_t(\bs)$ in $\log$-scale (top right); the diffeomorphism warp $w(\bs)$ (bottom left); the warped and smoothed forecast in $\log$-scale, $\tilde{X}_t(w(\bs))$, used to generate the data. The concentrations are measured in $\mu g/m^3$.}
\label{fig:sim_1}
\end{figure}

We consider four data generation processes. In the first case, data is generated by a simple linear regression (SLR) model with the forecast as the predictor, i.e., \[Y_t(\bs) = \beta_0 + \beta_1X_t(\bs) + \epsilon_t(\bs),\] with $\epsilon_t(\bs) \iid N(0,1)$, $\beta_0 = 1.5$ and $\beta_1 = 0.25$. Second, we use the smoothed forecast predictor \[Y_t(\bs) = \beta_0 + \sum_{l=1}^L\beta_l\tilde{X}_{lt}(\bs) + \epsilon_t(\bs),\] where $L=10$, $\epsilon_t(\bs) \iid N(0,1)$, $\beta_0 = 1.5$ and $\beta_l$ were decreasingly ordered realizations of a $N(0.25,0.0625)$ random variate for $l = 1,2, \ldots, 10$. The descending order of the coefficients ensures that the low frequency terms have higher weights than high frequency terms. The final two cases have a warped and smoothed forecast as predictor \[Y_t(\bs) = \beta_0 + \sum_{l=1}^L\beta_l\tilde{X}_{lt}(\tilde{w}(\bs)) + \epsilon_t(\bs)\] with $\tilde{w}(\cdot)$ being a warping function and the remaining components of the model being the same as in the previous scenario. These final two cases are distinguished by their warping function. The first warping function is the translation warp \[w(\bs) = \bs + \begin{pmatrix}
0.16\\
0.16 
\end{pmatrix}.\] The second warping function we used was diffeomorphism warp~\citep{guan2018computer} that preserves the boundaries of the image. For $0 \leq s_1, s_2 \leq 1$, \begin{align*}
    w_1(\bs) & = s_1 - 2\theta_1 s_2 \sin s_1 \cos s_2 (\cos \pi s_1 + 1)(\cos \pi s_2 + 1)\\
    w_2(\bs) & = s_2 - 2\theta_2 s_1 \sin s_1 \sin s_2 \cos \frac{\pi s_1}{2} \cos \frac{3\pi s_2}{2},
\end{align*} where $\theta_1$ and $\theta_2$ are tuning parameters jointly deciding the location, direction and extent of the warp set equal to $\theta_1 = 0.1$ and $\theta_2 = 0.5$. A visualization of the data generation process can be seen in Figure \ref{fig:sim_1}, including the original forecast, smoothed forecast, the diffeomorphism warp applied to the smoothed forecast and the synthetic forecast, warped and smoothed from the original. To investigate the effect of number of monitor stations, we select $25$, $50$ or $100$ monitor station observations randomly on the grid for each of the data generation processes.

For each of these scenarios, we fit a simple linear model to the data as well as three versions of the model proposed in Section \ref{s:model} with or without the warping and smoothing components. For each method, a Markov Chain Monte Carlo (MCMC) chain was run for $20,000$ iterations, of which the first $10,000$ iterations were discarded as burn-in samples.

To compare models we use mean squared error (MSE) and mean absolute deviation (MAD) computed using the posterior mean as point forecast and pointwise coverage of $95\%$ intervals and continuous ranked probability score (CRPS). The posterior predictive densities for the warped outputs can be skewed, heavy-tailed or even multi-modal and so metrics based on point predictions (MSE or MAD) may not capture the uncertainty properly. To evaluate the entire predictive distribution, CRPS \citep{gneiting2007strictly} is therefore a more meaningful choice since it is a measure of integrated squared difference between the cumulative distribution (CDF) function of the forecast and the corresponding CDF of the observations.

\begin{figure}
\begin{subfigure}{.49\textwidth}
    \centering
    \includegraphics[height = 0.645\textwidth]{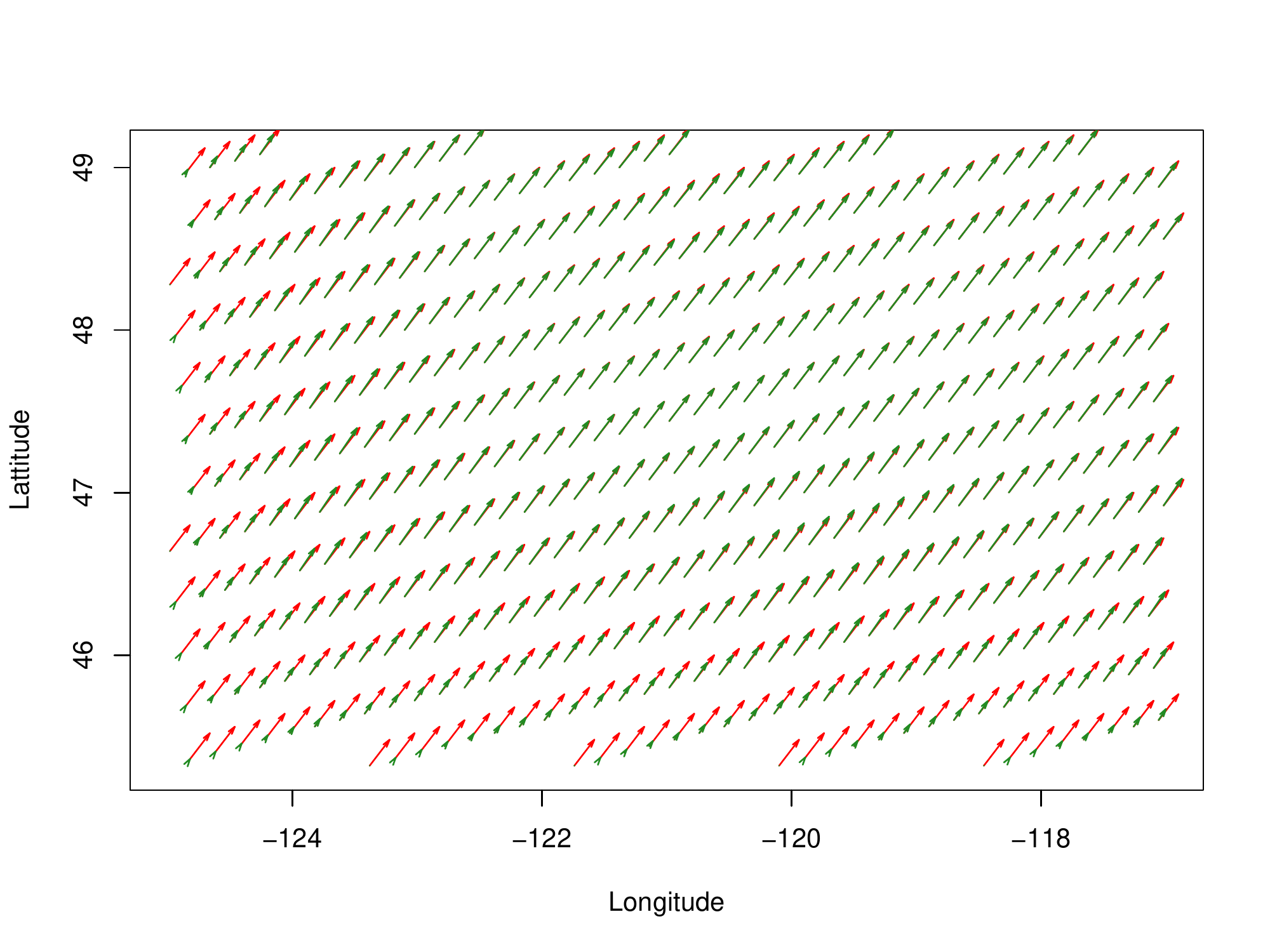}
\end{subfigure}
\begin{subfigure}{.49\textwidth}
    \centering
    \includegraphics[height = 0.645\textwidth]{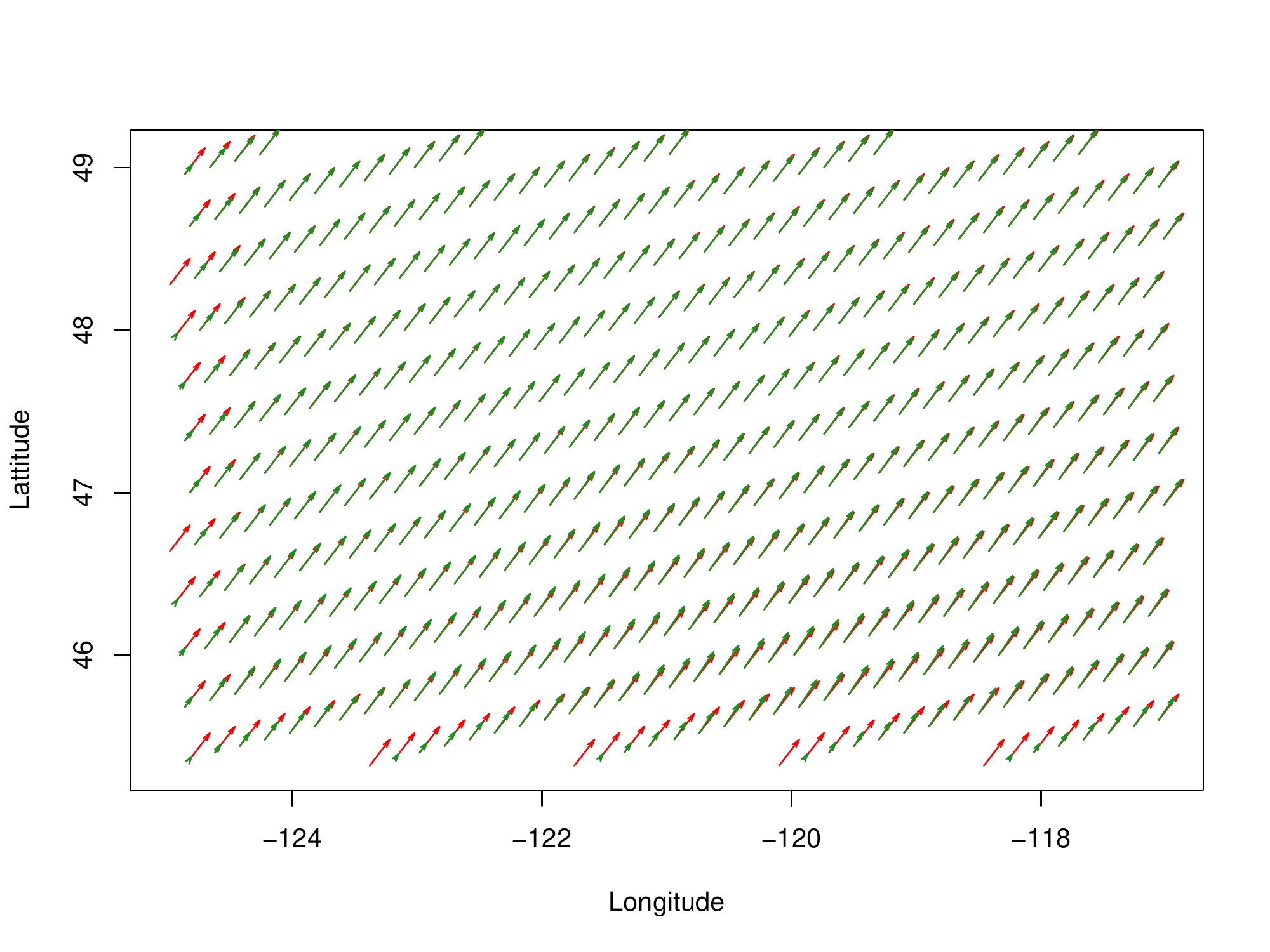}
\end{subfigure}
\begin{subfigure}{.49\textwidth}
    \centering
    \includegraphics[height = 0.645\textwidth]{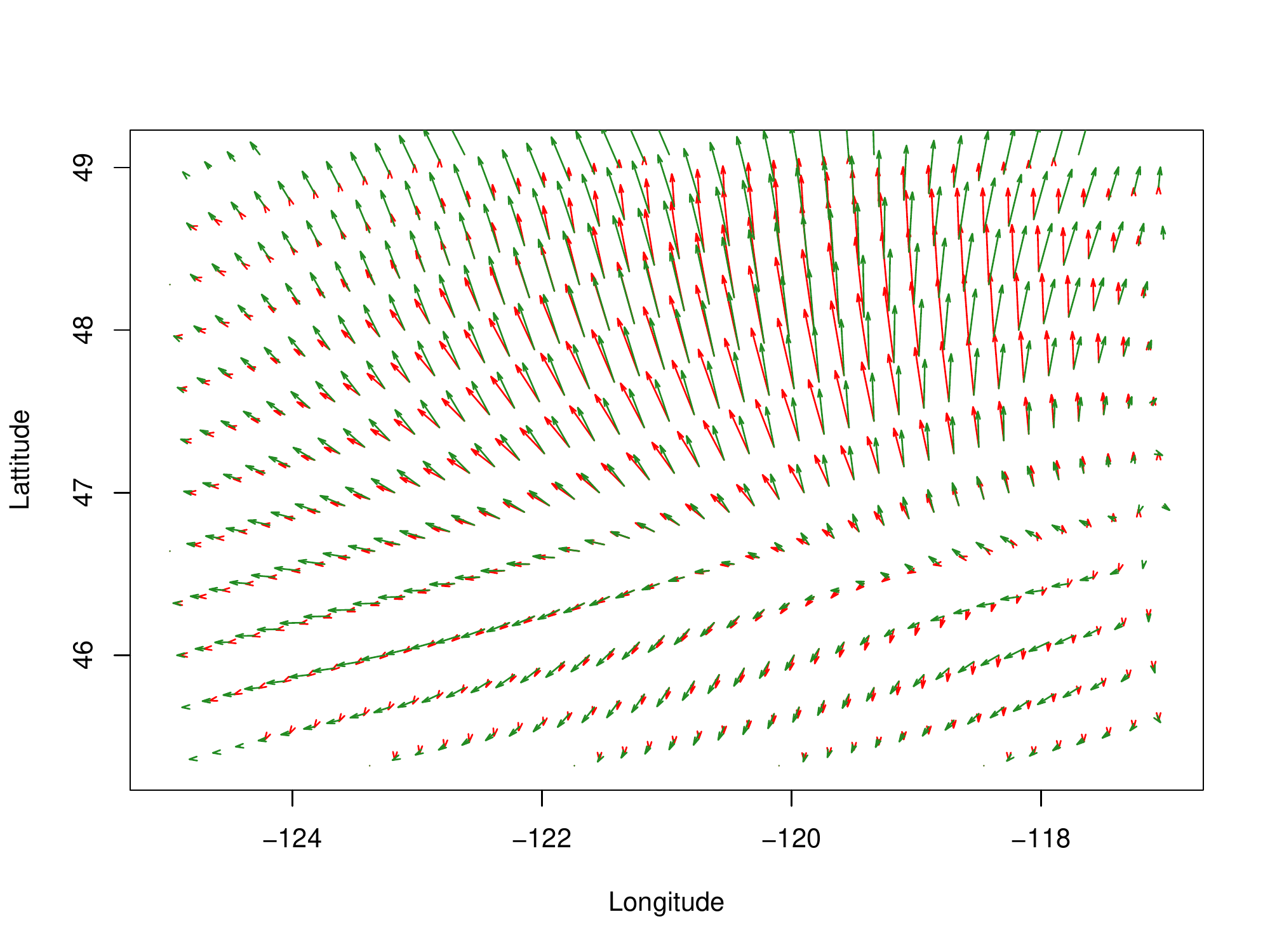}
\end{subfigure}
\begin{subfigure}{.49\textwidth}
    \centering
    \includegraphics[height = 0.645\textwidth]{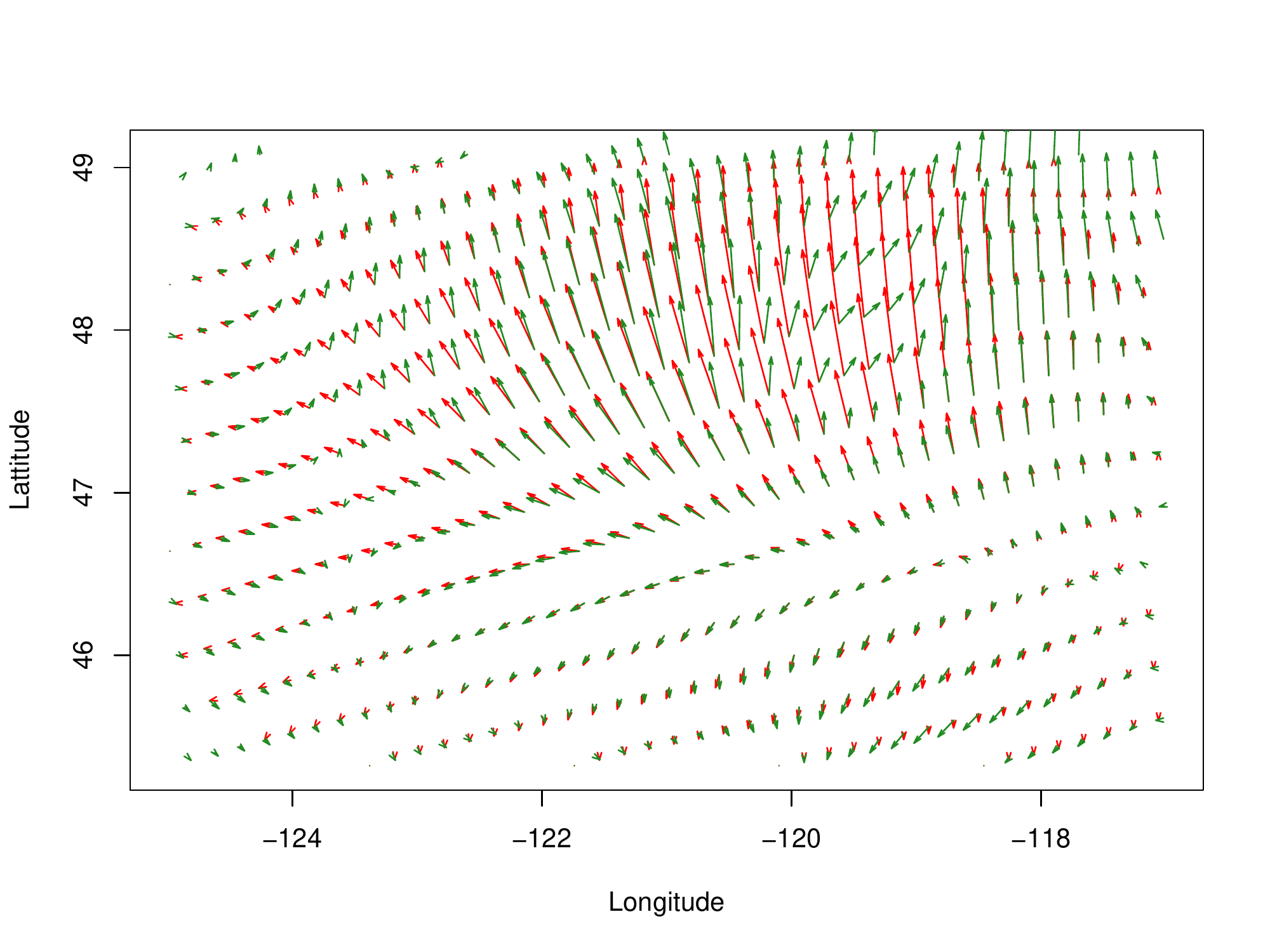}
\end{subfigure}
\caption{True (red) and estimated (green) warps for the translation warp (top row) and diffeomorphism warp (bottom row) for simulated data with $n= 25$ (left) and $n = 100$ (right).}
\label{fig:sim_2}
\end{figure}

We compute the $3$-day ahead forecast and compute the MSE, MAD, coverage and CRPS for the forecast of $n$ monitor stations. For each of the $12$ scenarios and for the corresponding $30$ datasets in each scenario, MSE, MAD, coverage and CRPS for each of the four models are computed and averaged over space and time for all datasets. The MSE (MAD is similar) and CRPS for these cases are reported in Tables \ref{tab:sim_1} and \ref{tab:sim_2}. For all methods and cases, coverage is always over $95\%$ and so it is not reported. Figure \ref{fig:sim_2} presents a comparison of the true and estimated (posterior mean) of warping function $w(\bs)$ for data sets with $n=25$ and $n=100$ from scenarios $3$ and $4$.

\begin{table}[!h]
\setlength{\tabcolsep}{0 pt}
    \centering
    \begin{tabular}{>{\small}c@{\hspace{0 pt}} >{\small}c@{\hspace{1 pt}} >{$}c<{$}@{\hspace{7 pt}} c@{\hspace{7 pt}} c@{\hspace{7 pt}} c@{\hspace{7 pt}} c}
    \hline
         Warp&Smoothing&n& SLR& \multicolumn{3}{c}{Proposed Model}\\
         \cline{5-7}
         & & & &Smooth&Warp& Both\\
         \hline
    &  & 25 & \textbf{1.01(0.05)} & 1.02(0.05) & 1.02(0.05) & 1.03(0.05) \\ 
   None & None & 50 & \textbf{1.00(0.03)} & \textbf{1.00(0.03)} & \textbf{1.00(0.03)} & 1.01(0.03) \\ 
    &  & 100 & \textbf{1.00(0.02)} & \textbf{1.00(0.02)} & \textbf{1.00(0.02)} & \textbf{1.00(0.02)} \\ 
    \hline 
    &  & 25 & 1.27(0.06) & \textbf{1.02(0.05)} & 1.35(0.06) & 1.03(0.05) \\ 
   None & Spectral & 50 & 1.30(0.03) & \textbf{1.00(0.03)} & 1.33(0.04) & 1.01(0.03) \\ 
    &  & 100 & 1.31(0.02) & \textbf{1.00(0.02)} & 1.35(0.02) & \textbf{1.00(0.02)} \\ 
  \hline
    &  & 25 & 1.69(0.08) & 1.34(0.06) & 1.60(0.08) & \textbf{1.19(0.13)} \\ 
   Translation & Spectral & 50 & 1.62(0.04) & 1.25(0.04) & 1.45(0.05) & \textbf{1.07(0.08)} \\ 
    &  & 100 & 1.64(0.03) & 1.25(0.02) & 1.49(0.04) & \textbf{1.09(0.10)} \\ 
  \hline
    &  & 25 & 1.53(0.07) & 1.32(0.06) & 1.46(0.07) & \textbf{1.17(0.09)} \\ 
   Diffeomorphism & Spectral & 50 & 1.50(0.03) & 1.23(0.03) & 1.39(0.04) & \textbf{1.10(0.04)} \\ 
    &  & 100 & 1.56(0.03) & 1.26(0.02) & 1.49(0.03) & \textbf{1.12(0.05)} \\ 
   \hline
    \end{tabular}
    \caption{MSE (standard error) estimates (in $\mu g^2/m^6$) for the proposed model with both smoothing and warping components, only smoothing component and only warping component along with a SLR model for different scenarios. The lowest MSE value in each case is in bold.}
    \label{tab:sim_1}
\end{table}

\begin{table}[!h]
\setlength{\tabcolsep}{0 pt}
    \centering
    \begin{tabular}{>{\small}c@{\hspace{0 pt}} >{\small}c@{\hspace{1 pt}} >{$}c<{$}@{\hspace{7 pt}} c@{\hspace{7 pt}} c@{\hspace{7 pt}} c@{\hspace{7 pt}} c}
    \hline
         Warp&Smoothing&n& SLR& \multicolumn{3}{c}{Proposed Model}\\
         \cline{5-7}
         & & & &Smooth&Warp& Both\\
         \hline

  &  & 25 & 0.78(0.02) & 0.78(0.02) & 0.78(0.02) & \textbf{0.77(0.02)} \\ 
   None & None & 50 & 0.78(0.01) & \textbf{0.77(0.01)} & \textbf{0.77(0.01)} & \textbf{0.77(0.01)} \\ 
    &  & 100 & 0.79(0.01) & 0.79(0.01) & 0.79(0.01) & \textbf{0.78(0.01)} \\ 
  \hline
    &  & 25 & 0.87(0.02) & \textbf{0.78(0.02)} & 0.85(0.02) & \textbf{0.78(0.02)} \\ 
   None & Spectral & 50 & 0.89(0.01) & \textbf{0.77(0.01)} & 0.85(0.01) & \textbf{0.77(0.01)} \\ 
    &  & 100 & 0.90(0.01) & \textbf{0.79(0.01)} & 0.86(0.01) & \textbf{0.79(0.01)} \\ 
  \hline
    &  & 25 & 1.00(0.02) & 0.85(0.02) & 0.88(0.03) & \textbf{0.78(0.04)} \\ 
   Translation & Spectral & 50 & 0.99(0.01) & 0.82(0.01) & 0.84(0.02) & \textbf{0.77(0.01)} \\ 
    &  & 100 & 1.00(0.01) & 0.83(0.01) & 0.90(0.01) & \textbf{0.78(0.01)} \\ 
  \hline
    &  & 25 & 0.96(0.03) & 0.84(0.02) & 0.85(0.03) & \textbf{0.76(0.03)} \\ 
   Diffeomorphism & Spectral & 50 & 0.95(0.01) & 0.82(0.01) & 0.83(0.01) & \textbf{0.75(0.02)} \\ 
    &  & 100 & 0.97(0.01) & 0.83(0.01) & 0.90(0.02) & \textbf{0.77(0.01)} \\ 
   \hline
    \end{tabular}
    \caption{CRPS (standard error) estimates for the proposed model with both smoothing and warping components, only smoothing component and only warping component along with an OLS model for different scenarios. The lowest CRPS value in each case is in bold.}
    \label{tab:sim_2}
\end{table}

From Tables \ref{tab:sim_1} and \ref{tab:sim_2}, all methods perform similarly when data are generated from the SLR model. Therefore the added complexities of the full model do not result in overfitting in this case. The full model has smaller MSE and CRPS than the SLR model in the second case. Although the model with only smoothing component is somewhat better as that matches the true data generation model. In the later two cases, the full model provides the best results. The performance of all methods improve with increasing values of $n$. This is reflected in Figure \ref{fig:sim_2} which compares the true and estimated warp for both the warps used in this study. In both cases, the estimates are closer to the true value for $n=100$ than for $n=25$. The estimation is more accurate for the translation warp, compared to the more complicated diffeomorphism warp.

\section{Application to $\PM$ Forecasting in Washington State}\label{s:data}


In our simulation study, we fix the warping function to be constant over time. However, for the wildland fire application, the warp likely varies over time, following changes in the location of the fires and wind field. Therefore, we analyze the data separately by day with the first $18$ hours of data as training and forecast on the next $6$ hours for each of the $35$ days. This strikes a balance between flexibility to capture dynamics of the warping function while still providing sufficient training data to estimate the warping function. The priors, models, computational details and metric of comparison are the same as the simulation study.

\begin{figure}
    \begin{subfigure}{0.49\textwidth}
    \centering
    \includegraphics[height = 0.3\textheight]{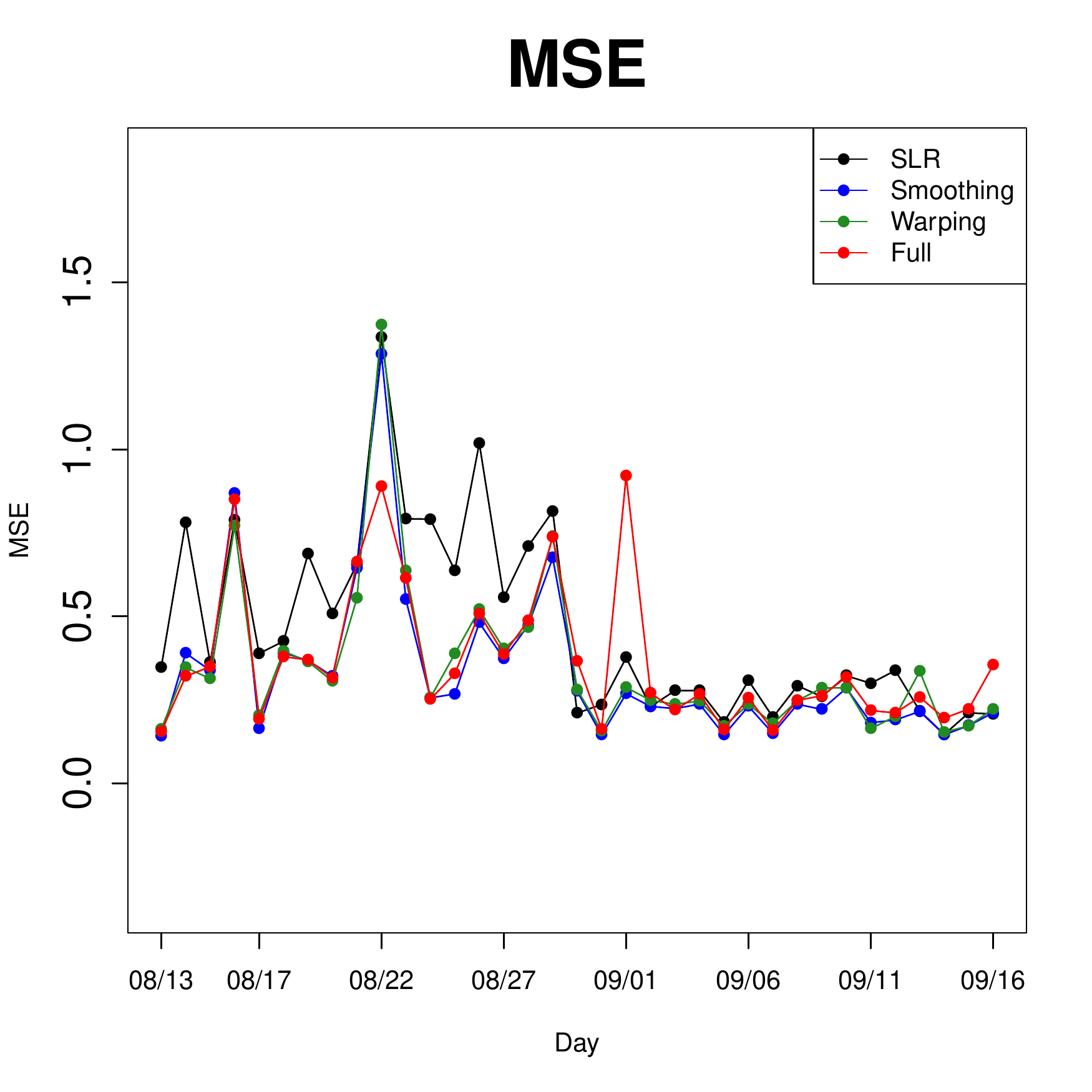}
\end{subfigure}
\begin{subfigure}{0.49\textwidth}
    \centering
    \includegraphics[height = 0.3\textheight]{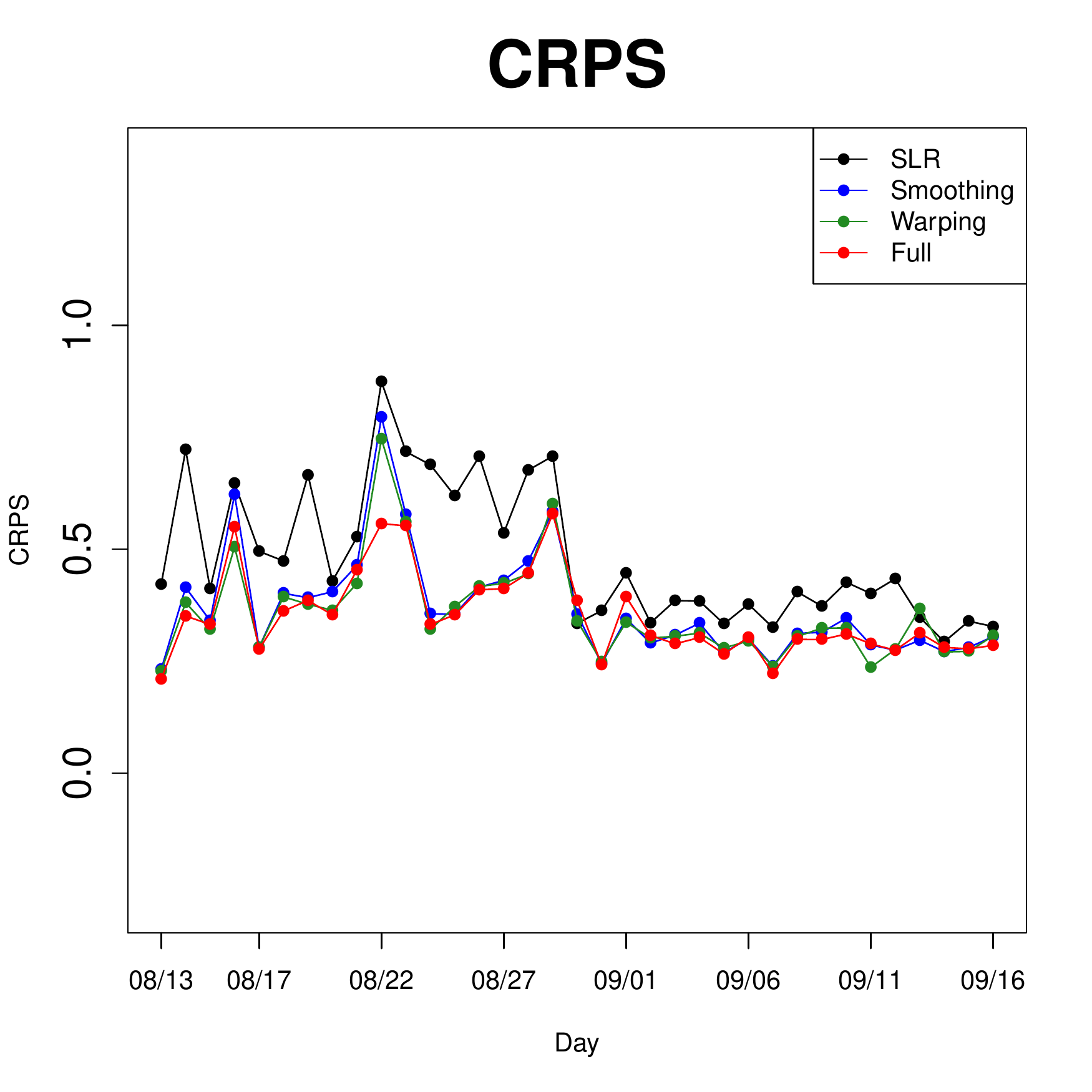}
\end{subfigure}
    \caption{Daily prediction MSE (left panel) in $\mu g^2/m^6$ and CRPS (right panel) for the four models}
    \label{fig:dat_mse_crps}
\end{figure}

For each day, we compute predictive MSE, MAD, coverage and CRPS averaged over space and time. These metrics, averaged over days, are presented in Table \ref{tab: data_1}. Day-by-day comparisons for MSE and CRPS are presented in Figure \ref{fig:dat_mse_crps} (similar figure for MAD and coverage are available in \href{subsec: dat_fig_extra}{Appendix}).

\begin{table}[ht]
\centering
\begin{tabular}{ccccc}
  \hline
 & SLR & Smoothing & Warp & Full \\ 
  \hline
MSE & 0.4624 & 0.3357 & 0.3517 & 0.3675 \\ 
  MAD & 0.4983 & 0.4081 & 0.4198 & 0.4219 \\ 
  Coverage & 0.9237 & 0.9294 & 0.9126 & 0.9019 \\ 
  CRPS & 0.4842 & 0.3688 & 0.3570 & 0.3500 \\  
   \hline
\end{tabular}
\caption{Values of MSE (in $\mu g^2/m^6$), MAD (in $\mu g/m^3$), coverage and CRPS for the data analysis for the different methods, averaged over days.}
\label{tab: data_1}
\end{table}

The models with smoothing, warping or both perform significantly better than SLR. Smoothing leads to the largest reduction in MSE and MAD while the full model leads to the largest reduction in CRPS. Therefore, smoothing appears to be sufficient if only a point estimate is required, but including the warping function provides a better fit to the full predictive distribution.

\begin{figure}
\begin{subfigure}{.49\textwidth}
    \centering
    \includegraphics[width=\textwidth]{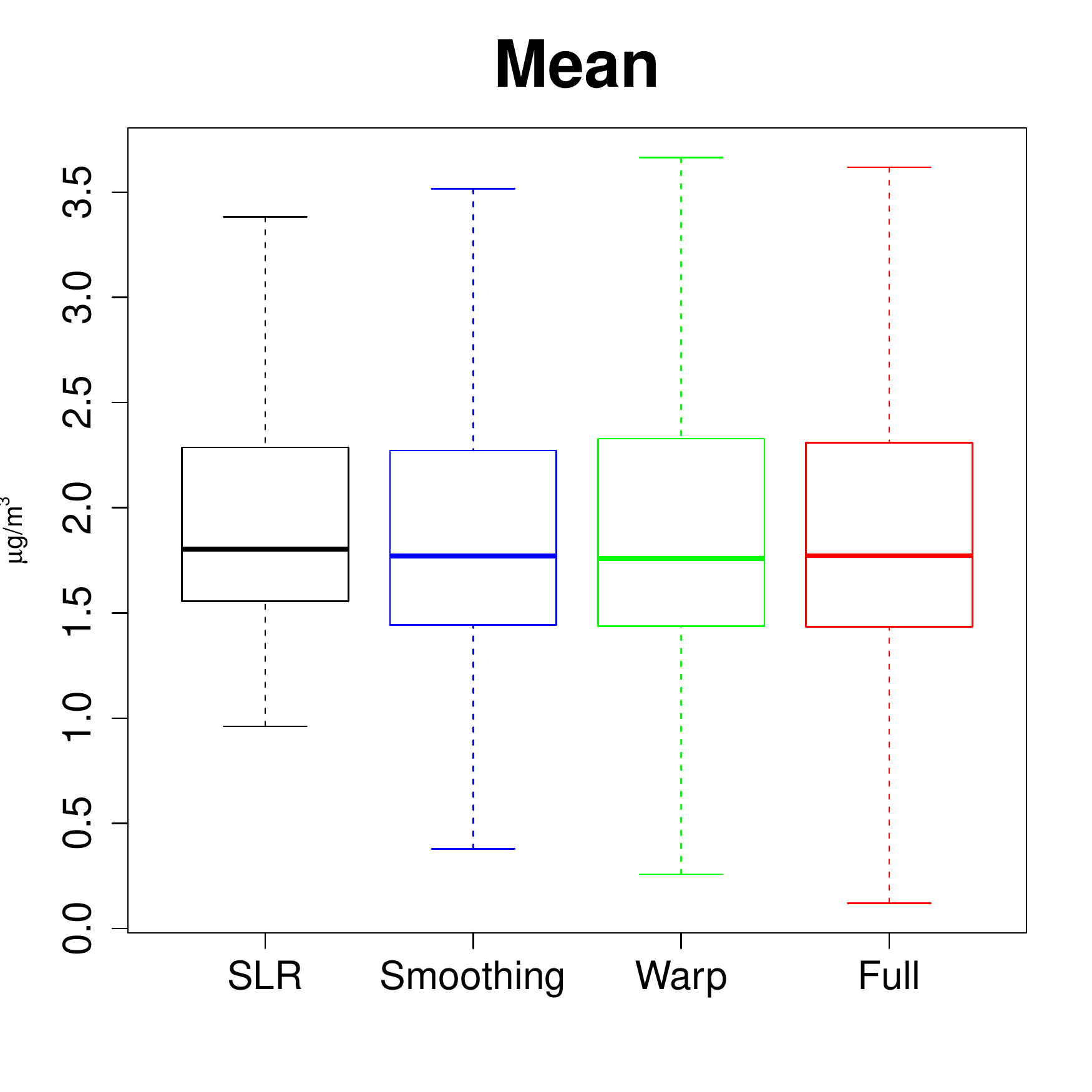}
\end{subfigure}
\begin{subfigure}{.49\textwidth}
    \centering
    \includegraphics[width=\textwidth]{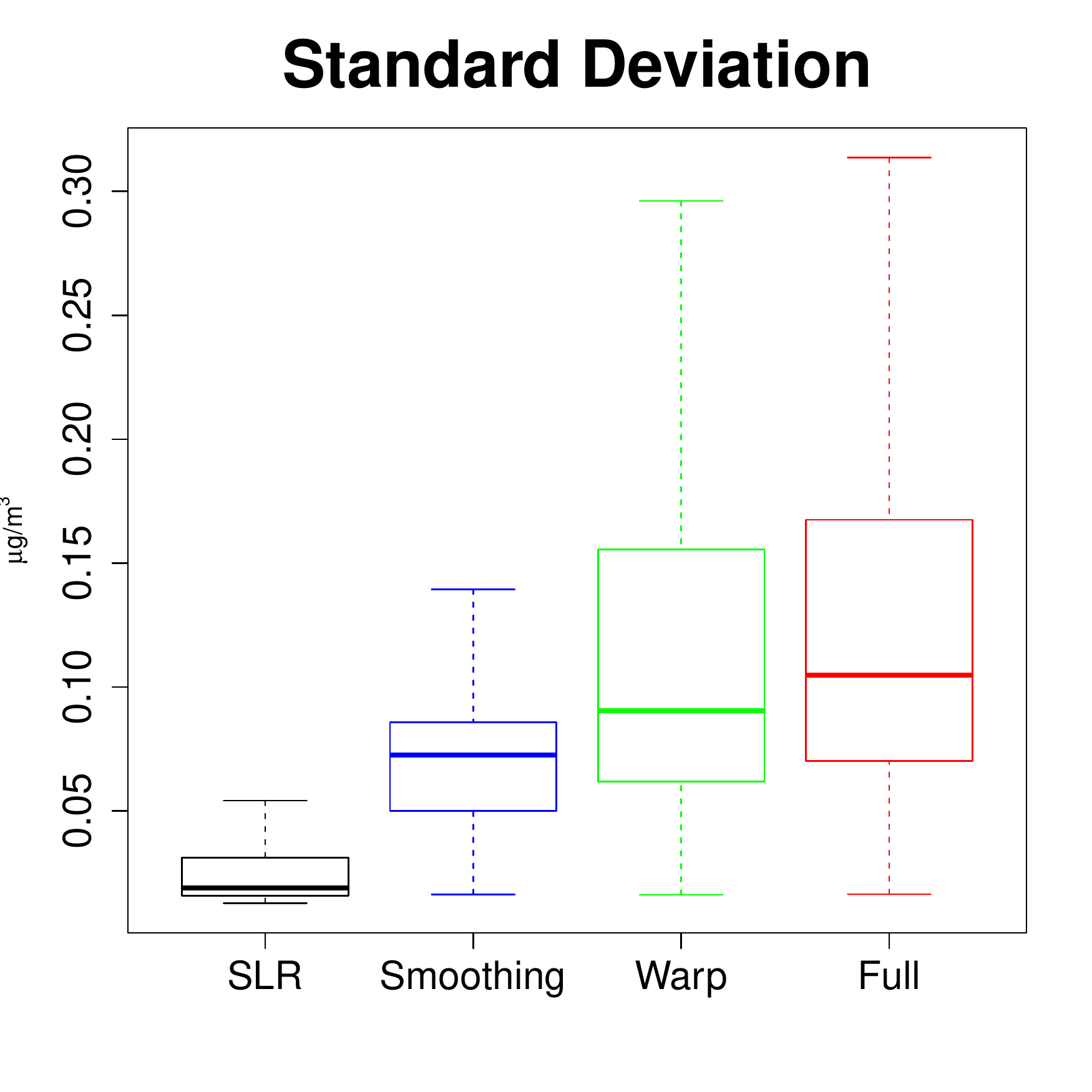}
\end{subfigure}
\begin{subfigure}{.49\textwidth}
    \centering
    \includegraphics[width=\textwidth]{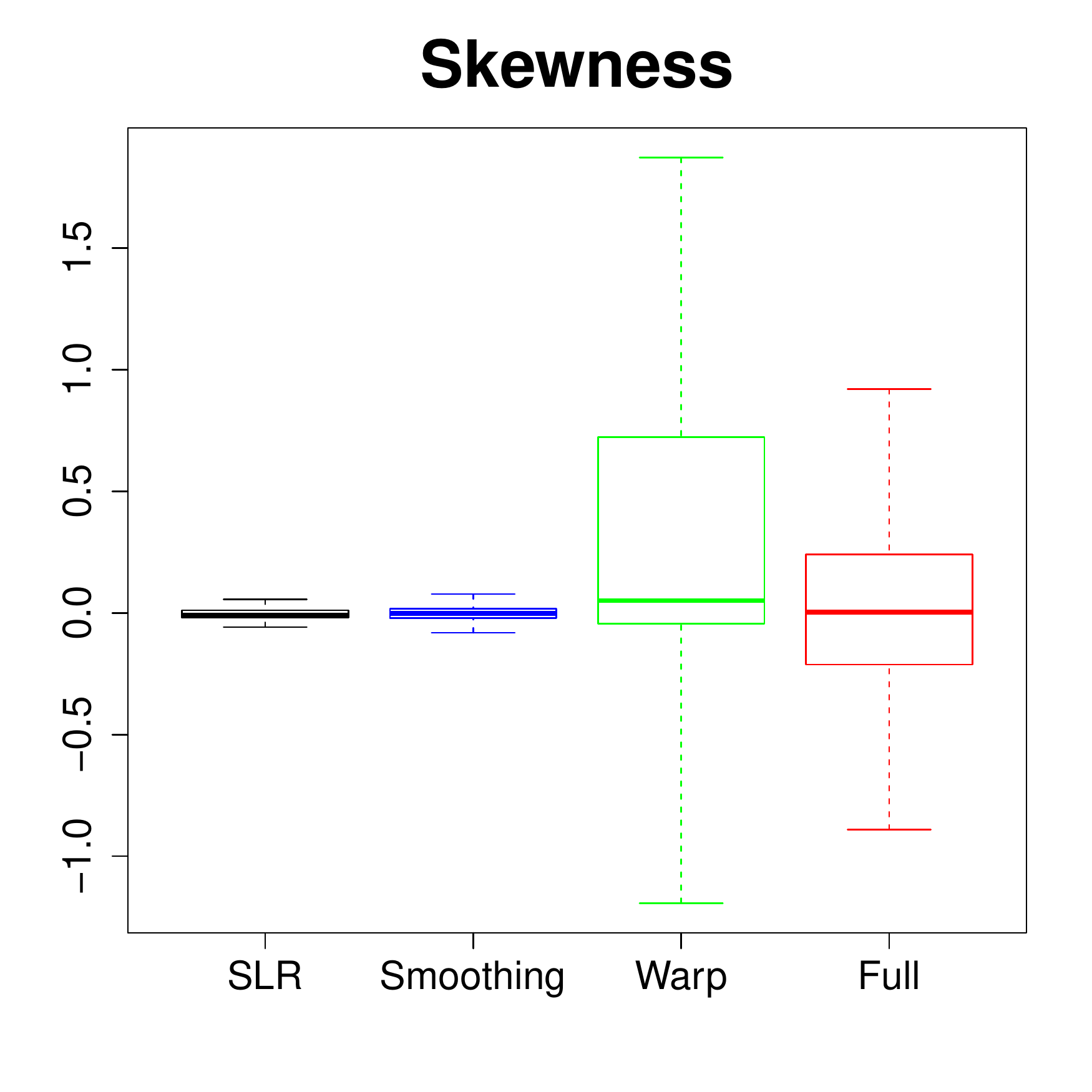}
\end{subfigure}
\begin{subfigure}{.49\textwidth}
    \centering
    \includegraphics[width=\textwidth]{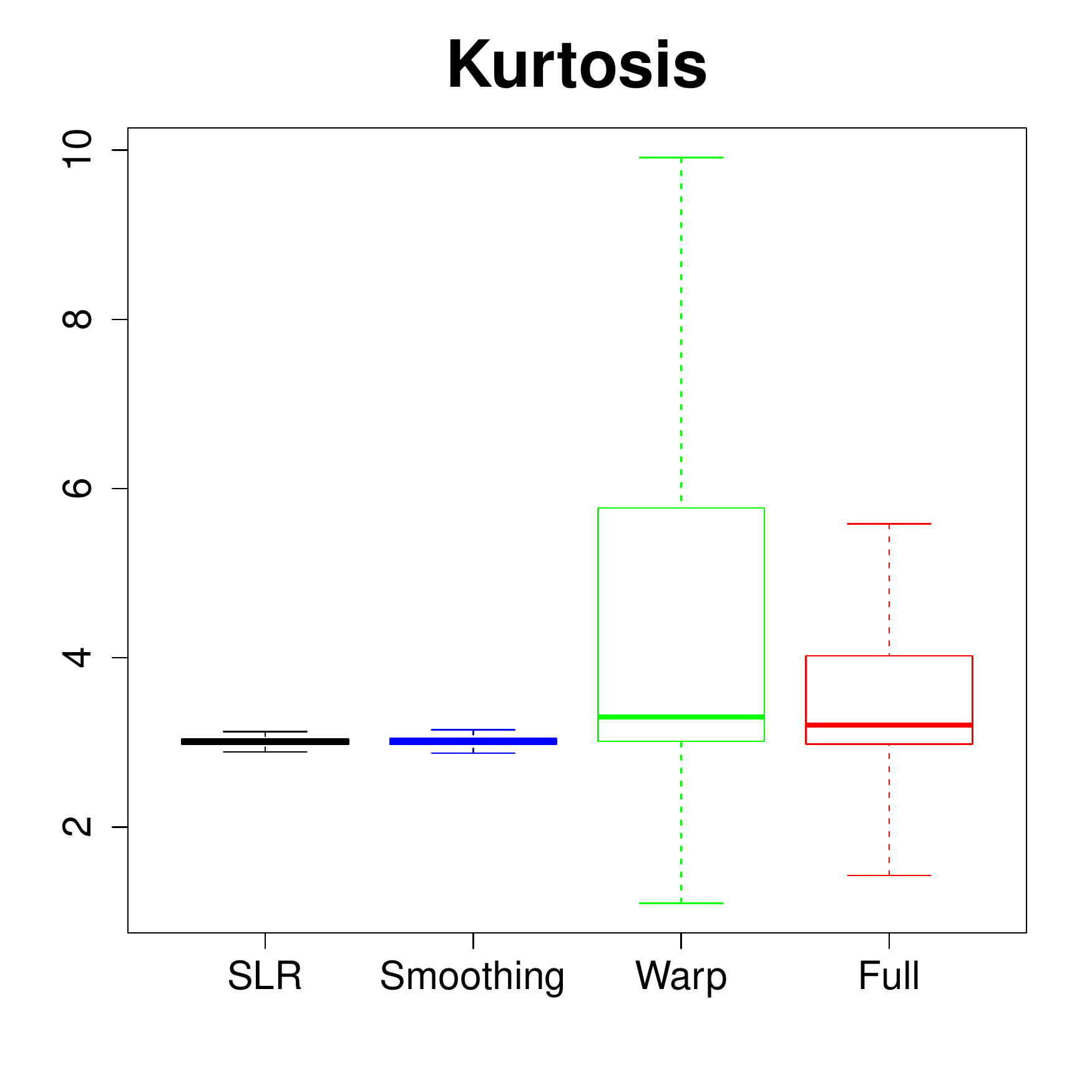}
\end{subfigure}
\caption{Boxplots of mean (in $\mu g/m^3$), standard deviation (in $\mu g/m^3$), skewness and kurtosis of the posterior predictive distributions for each model. The color schemes for each model is black (SLR), blue (Smoothing), green (Warp) and Red (Full) for each of the subfigures.}
\label{fig:data_box}
\end{figure}

To further illustrate how the warping models provide richer uncertainty quantification, we compute the posterior mean, standard deviation, skewness and kurtosis for each test set observation and present the distribution of these summary statistics as boxplots in Figure \ref{fig:data_box}. The mean values are similar for all models, but the warp based models exhibit higher skewness and kurtosis.

\begin{figure}
    \begin{subfigure}{0.49\textwidth}
        \centering
    \includegraphics[width=\textwidth]{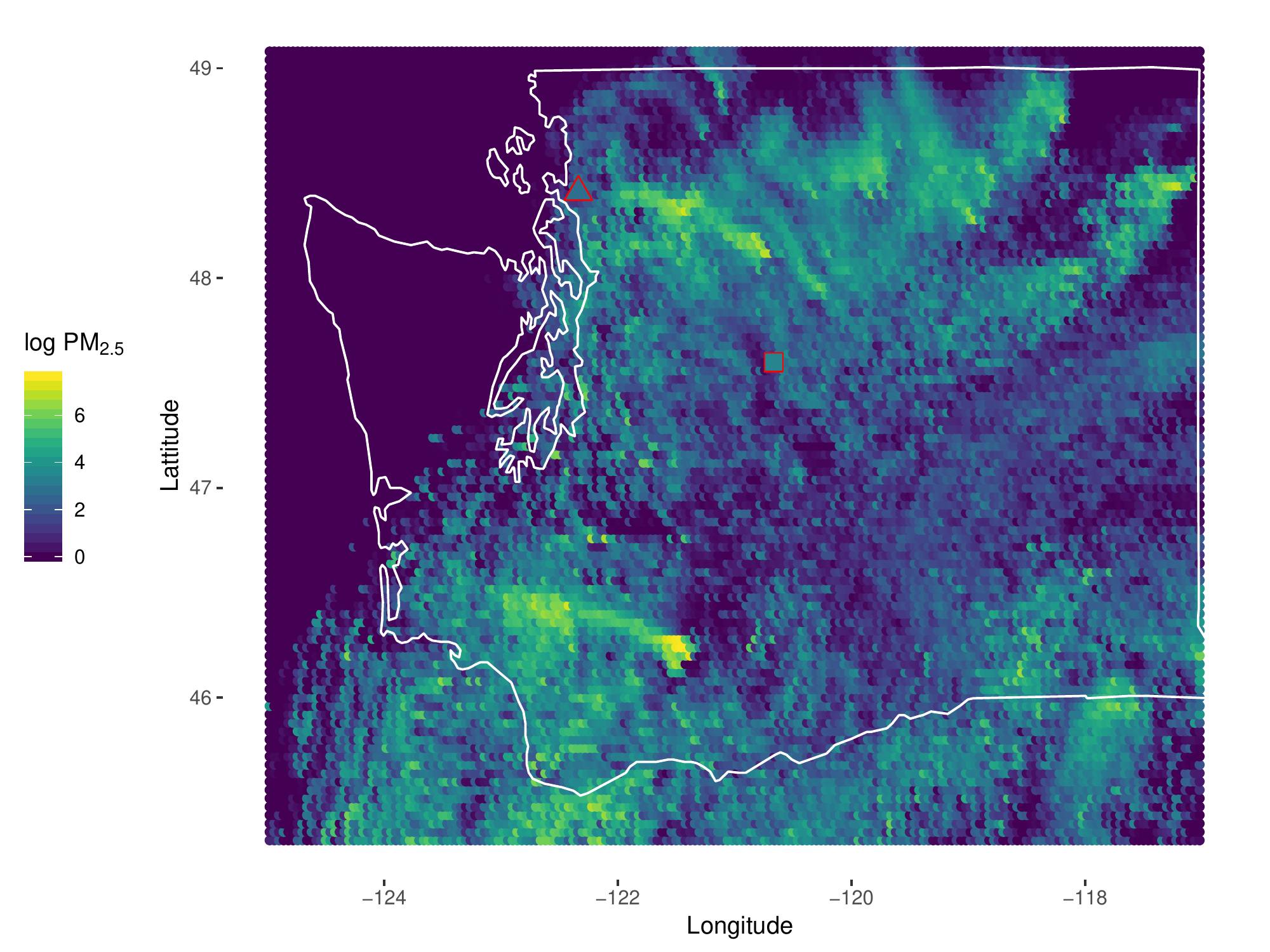}
    \end{subfigure}
    \begin{subfigure}{0.49\textwidth}
        \centering
    \includegraphics[width=\textwidth]{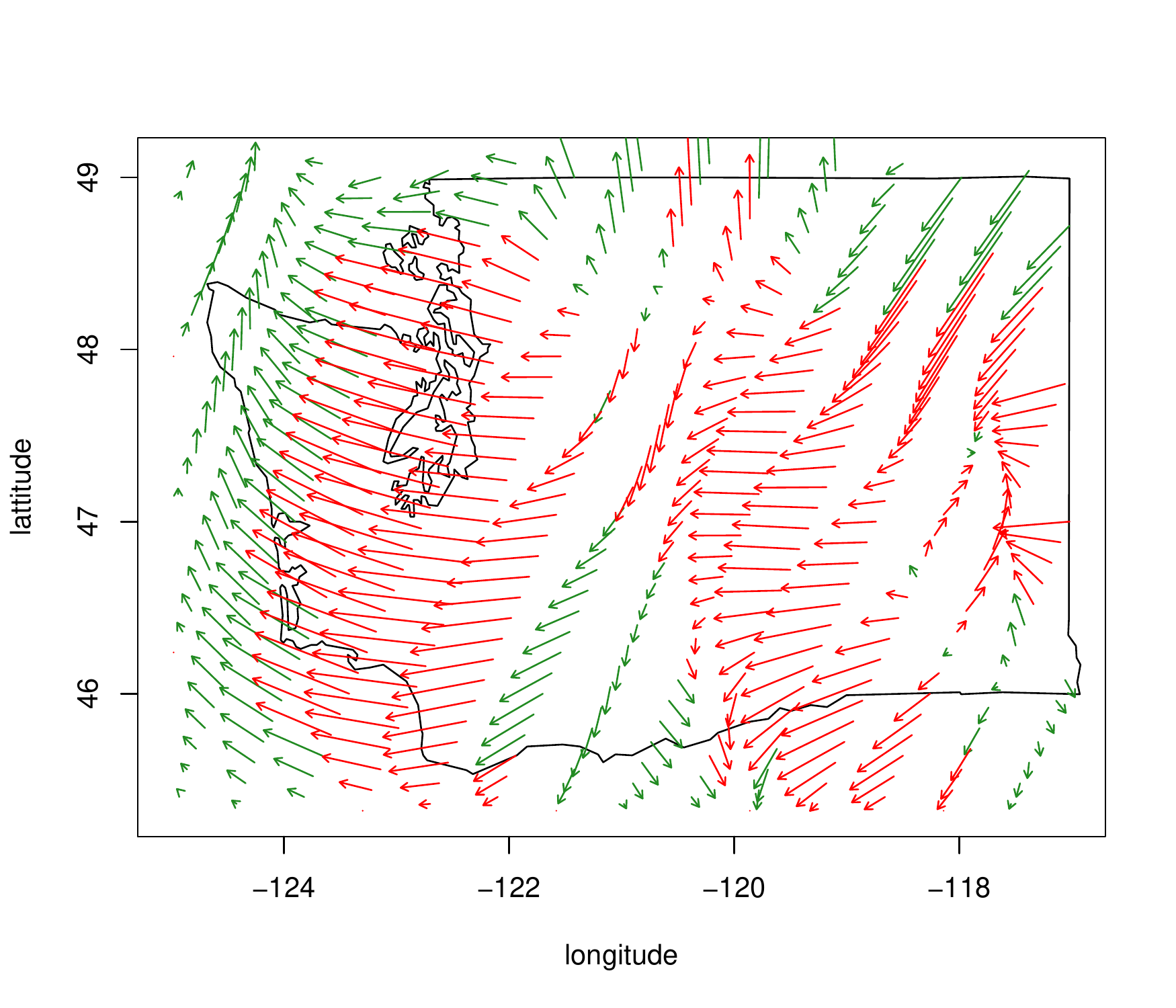}
    \end{subfigure}
    \begin{subfigure}{0.49\textwidth}
        \centering
    \includegraphics[width = \textwidth]{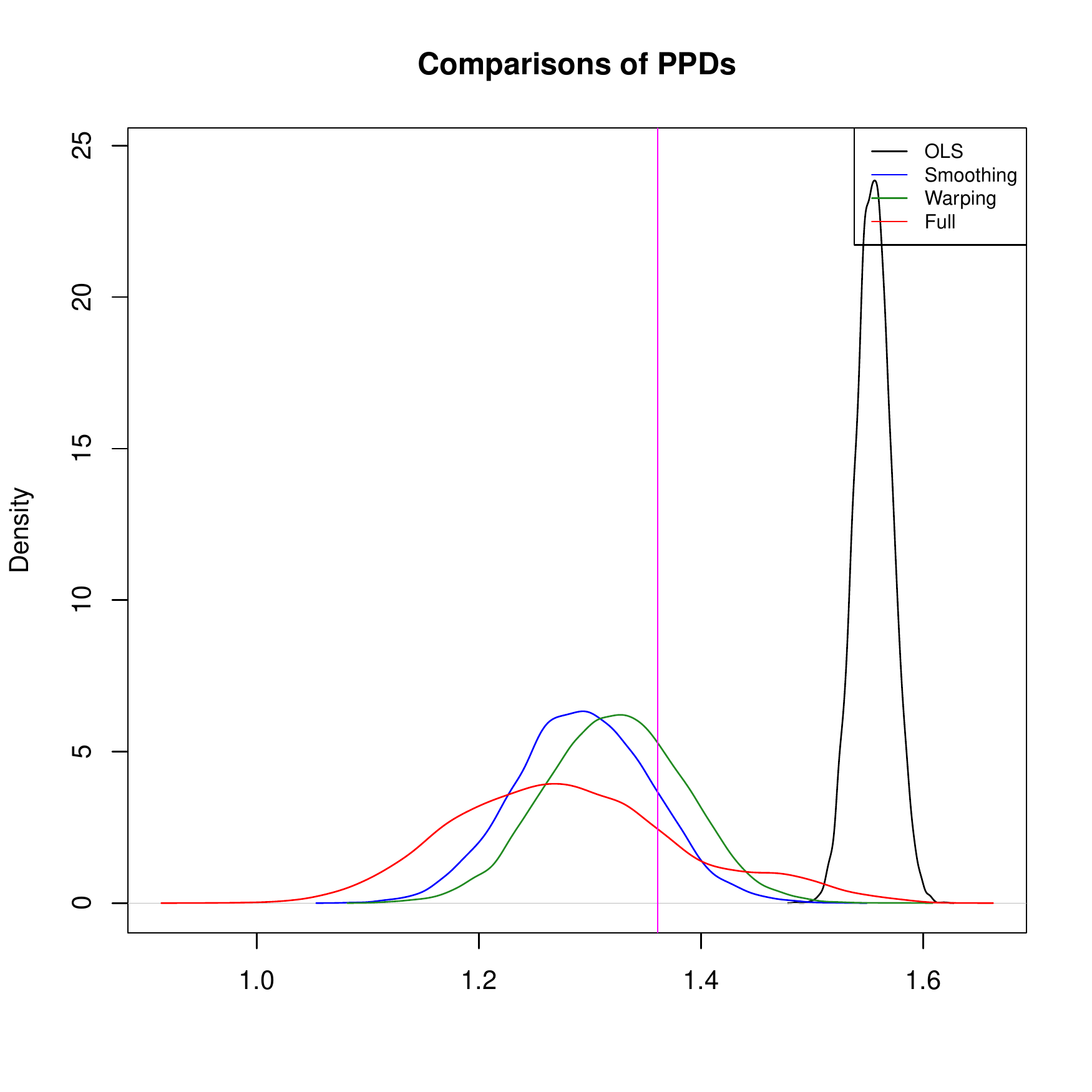}
    \end{subfigure}
    \begin{subfigure}{0.49\textwidth}
        \centering
    \includegraphics[width=\textwidth]{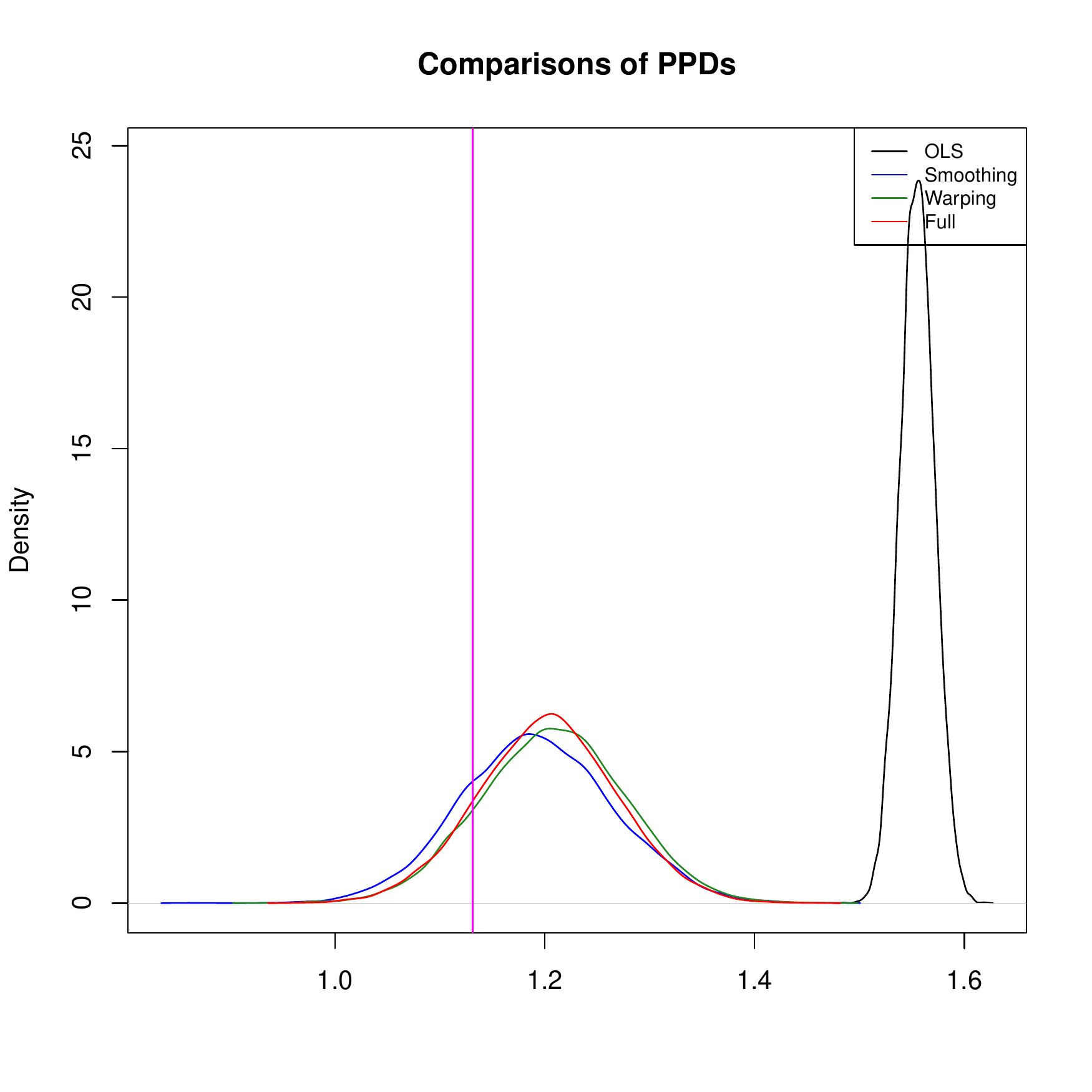}
    \end{subfigure}
    \caption{The numerical forecast of $\log$-concentration of $\PM (\mu g/m^3)$ on August 22, 2015 at 7:00 PM (top left) with two stations highlighted, one near the edge of fire (triangle) and one in the middle of fire (rectangle). The estimated warp for the day (top right) showing significant (red) and non-significant (green) warps. Comparisons of PPDs are made for the four competing models for the location marked as triangle (bottom left) and the location marked as rectangle (bottom right).}
\label{fig:data_ppd_comp}
\end{figure}

\begin{figure}
    \begin{subfigure}{0.49\textwidth}
        \centering
    \includegraphics[width=\textwidth]{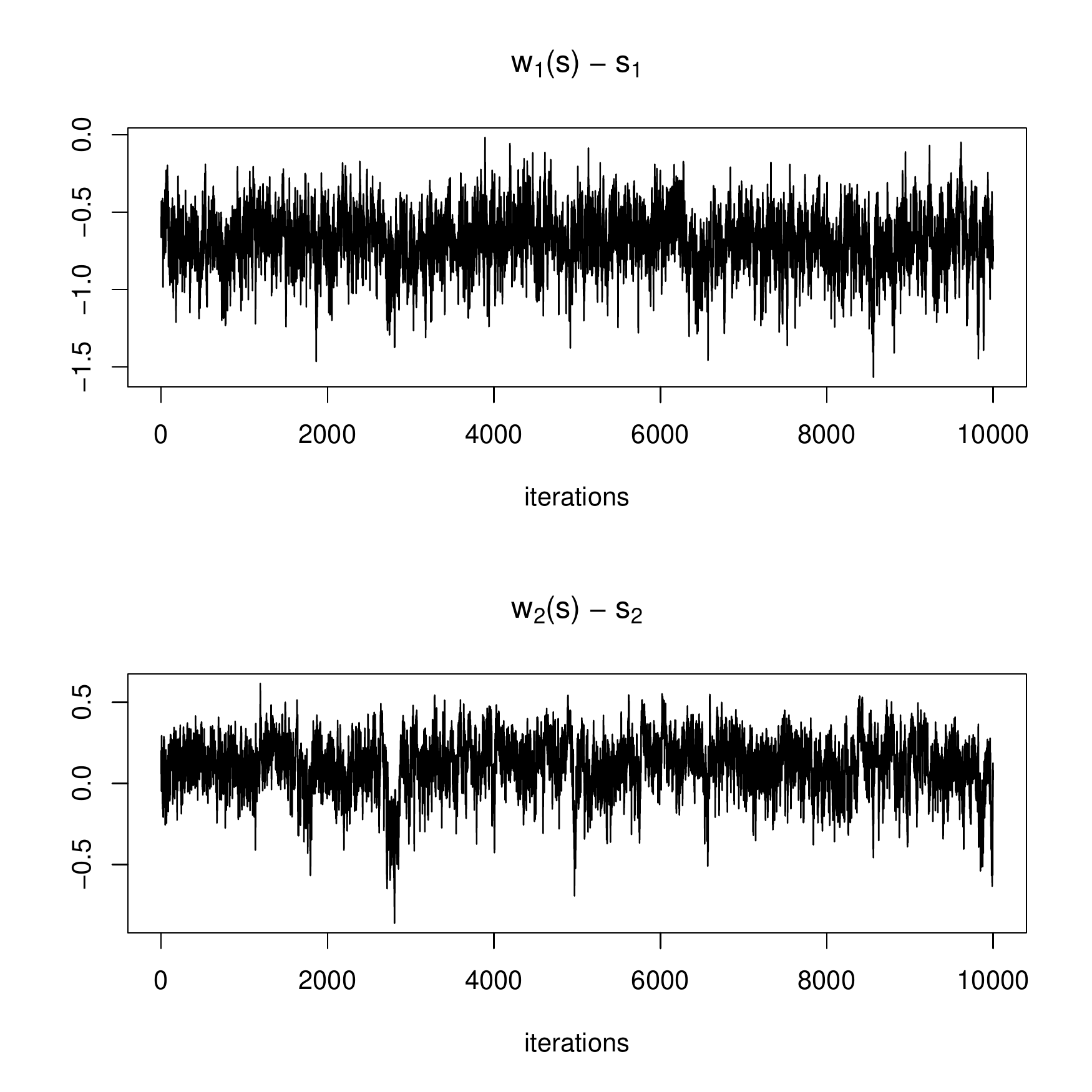}
    \end{subfigure}
    \begin{subfigure}{0.49\textwidth}
        \centering
    \includegraphics[width=\textwidth]{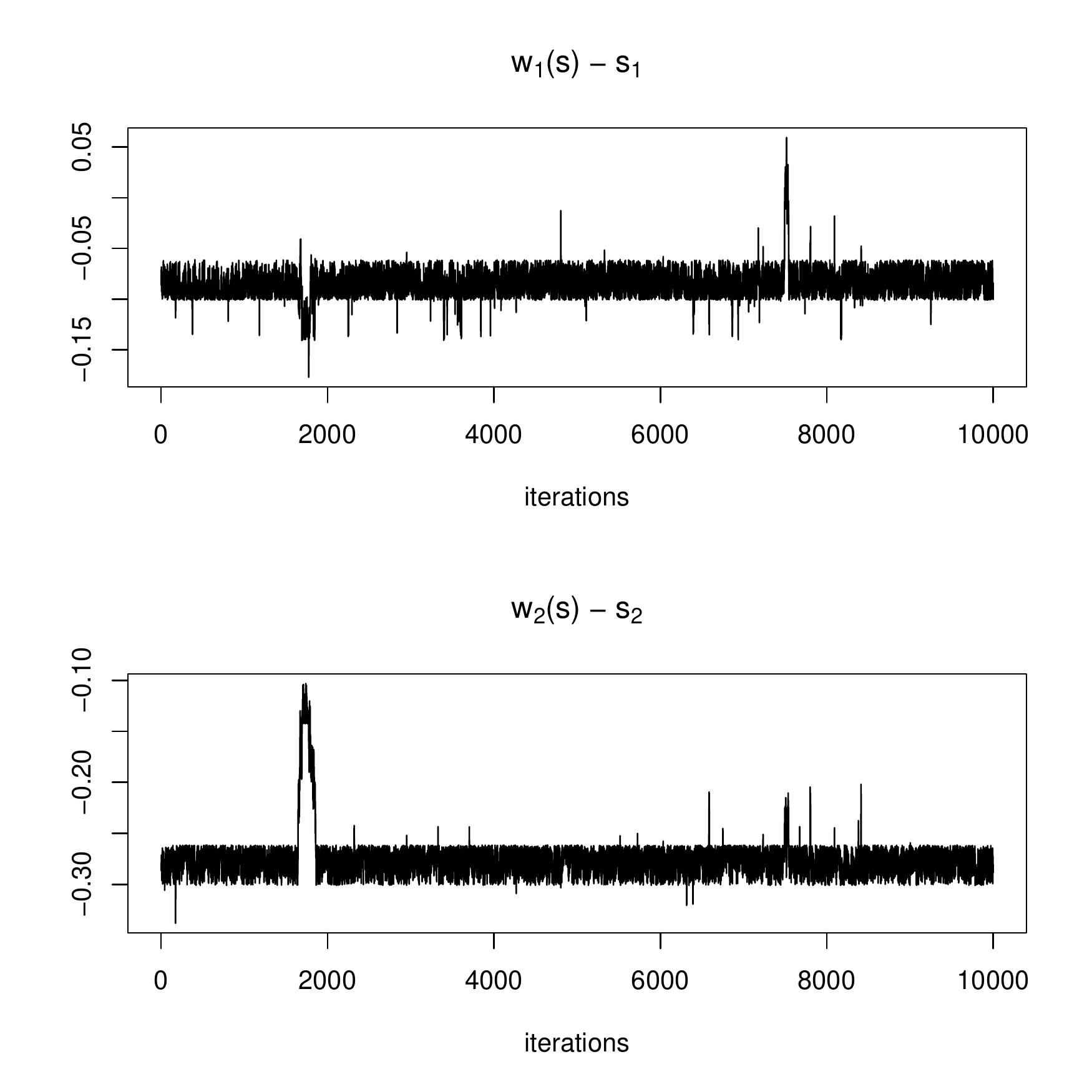}
    \end{subfigure}
    \caption{Trace plots for x (top) and y (bottom) coordinates of the estimated displacement due to warp ($w(\bs) - \bs$) for the two locations flagging in Figure~\ref{fig:data_ppd_comp}. Left panel is for the location at the edge of fire and right panel for the location in the middle of it.}
\label{fig:data_trace}
\end{figure}

Skewness and kurtosis often result from uncertainty in the warping function. For example, we compare the posterior predictive densities (PPD) of the four methods for a particular station located at the edge of the wildfire on August 22, 2015 at 7:00 PM in the left panel of Figure \ref{fig:data_ppd_comp}. One would expect high uncertainty in estimation for such a location. The PPD from SLR method misses the true value (magenta) by quite some margin, while the three methods capture the true value within their respective PPDs. The PPD for the full model estimator has a heavier right tail and smaller peak, indicating high variance and kurtosis capturing the uncertainty of estimation in such a location. On the other hand, comparing the densities for a location that is in the middle of the wildfire for the same day and time, shows that the PPDs behave similar to each other and have low skewness and thin tails, as can be seen in the right panel of Figure \ref{fig:data_ppd_comp}. Figure \ref{fig:data_ppd_comp} also shows the estimated warping function for the day. The red arrows imply a significant warp at the location at its base, while the green ones are non-significant (where warp at location $\bs$ is significant if the $95\%$ credible set of either component of $w(\bs) - \bs$ excludes zero). The trace plots for the estimate of  displacement due to the warping function ($w(\bs) - \bs$) for the two locations marked in Figure \ref{fig:data_ppd_comp} are presented in Figure \ref{fig:data_trace}. The location in the middle of the fire (right panel) has values around zero, meaning a non-significant warp at the location. The estimate for the location at the edge the fire is has a jagged trace with values away from zero, indicating a significant warp. In both cases, the MCMC algorithm mixes well.

\section{Concluding Remarks}\label{s:disc}

Motivated by an wildland fire application, we develop a new downscaling method that incorporates spectral smoothing and image warping techniques into a single downscaling method which is shown to improve forecast distributions for simulated and real data.

The above method can be extended to incorporate temporally varying warping function and to have an error distribution that is spatiotemporally correlated. The extension is simple in nature but will add severe complexity to an already complex model. This would, in our experience, require a lot of data points to successfully estimate the warping function and the spatiotemorally correlated error structure. The model may also be extended to a spatial extremes scenario which would require a different error structure and its successful estimation.

\section*{Acknowledgements}

We would like to thank the United States Forest Services (USFS) for providing the data. The authors were partially supported by NSF DMS-1638521, NIH ES027892, DOI 14-1-04-9 and KAUST 3800.2. We are grateful for this support.

\appendix


\section{Technical Details for the Model}\label{s:app_tech}

\subsection{Construction of Basis Functions for Spectral Smoothing}\label{subsec: smooth basis}
As mentioned in Section \ref{s:model}, we smooth our forecast using a spectral smoothing approach proposed by \cite{reich2014spectral}. This process, using fast Fourier transform and inverse fast Fourier transform, breaks the original forecasts $X_t(\bs)$ into several layers $X_{lt}(\bs)$ by weighting them with basis functions $V_{l}(\bomega)$. Each of the $X_{lt(\bs)}$s contains information about phenomenons of different scales. We mentioned some restrictions on the basis functions to be used in Section \ref{s:model}.

A common choice for choosing this basis function is the Bernstein polynomial basis function, as suggested by \cite{reich2014spectral}. This approach assumes that the dependence of frequency $\bomega$ in constructing the basis functions is solely on the magnitude of the frequency $|| \bomega ||$. With this assumption, the basis functions can be written as \beq \label{eq: bernstein}
V_{l}(\bomega) = V_{l}(|| \bomega ||) = \begin{pmatrix} L-1 \\
l-1 \end{pmatrix} \Lp \frac{|| \bomega ||}{2\pi} \Rp^{l-1} \Lp 1 - \frac{|| \bomega ||}{2\pi} \Rp^{L-l},  \eeq for $l=1,2,\ldots, L$. This set up ensures that $\int B_{l}(\bomega) \,\ d\bomega = 1 \,,\ \forall l$.

However, such representation of $\tilde{X}_{lt}(\bs)$ may be subject to identifiability issues because of how the basis functions are defined in Equation (\ref{eq: bernstein}). To avoid such issues, we follow \cite{reich2014spectral} and define \beq \label{eq: alias}
\bdelta = \begin{cases}
\bomega \mbox{ if } ||\bomega || \leq || \bar{\bomega} || \\
\bar{\bomega} \mbox{ if } ||\bomega || > || \bar{\bomega} ||
\end{cases} \in [0,2\pi), \eeq where $\bar{\bomega} = [\bbi(\omega_1 > 0)(2\pi - \omega_1),\bbi(\omega_2 > 0)(2\pi - \omega_2)]^{\sf T}$. After this, we define our basis functions as $V_{l}(\bomega) = V_{l}(|| \bdelta ||)$. This ensures we avoid aliasing issues while retaining the other properties.

\subsection{Computing}\label{ss:comp}

The warping function is not completely identifiable, that is to say that for two different warping functions $w_1(\bs) \neq w_2(\bs)$, we may have the same warped output $X_t(w_1(\bs)) = X_t(w_2(\bs))$ for some $\bs$ and at some timepoint $t$. If the two warping function differ only on how they map points with zero values to other points with zero values, then it is not possible to distinguish them. Assuming that the forecast would be non-constant over any region is unrealistic as it is bound to have regions with zero values, in general. This is not necessary for us to have the warping function identifiable, but it does create problems with convergence as parameters can fluctuate between two sets of values both of which give the same warped output.

Another concern for convergence is the large number of parameters in the model. The smoothing coefficients needed to be marginalized to achieve convergence in the full model. Convergence of component models (smoothing-only or warping-only) is much quickly achieved compared to the full model scenario and usually require no tricks such as marginalization, although we used marginalization for them as well. We used the simple Metropolis within Gibbs algorithm to run our MCMC chains throughout. Metropolis-adjusted Langevin algorithm (MALA) or Hamiltonian Monte Carlo (HMC) methods may provide quicker convergence but would add much complexity to each iteration.

\section{Supplemental Tables and Figures}\label{s:app_tabnfig}

\subsection{Additional Tables for MAD and Coverage Estimates from the Simulation Study}\label{subsec: sim_tab_extra}

We present here additional tables obtained from the simulation study detailed in Section \ref{s:sim}. These tables show the performance of the four models, the OLS model, the full model (see Section \ref{s:model}) and the two sub-models, smoothing only and warping only model (see Section \ref{s:sim}), in four different data generation scenarios with three different values of $n$ for each of the four cases. The results obtained here are similar to those in Section \ref{s:sim}.

\begin{table}[!h]
\setlength{\tabcolsep}{0 pt}
    \centering
    \begin{tabular}{>{\small}c@{\hspace{0 pt}} >{\small}c@{\hspace{1 pt}} >{$}c<{$}@{\hspace{7 pt}} c@{\hspace{7 pt}} c@{\hspace{7 pt}} c@{\hspace{7 pt}} c}
    \hline
         Warp&Smoothing&n& SLR& \multicolumn{3}{c}{Proposed Model}\\
         \cline{5-7}
         & & & & Warp &Smooth& Both\\
         \hline
  &  & 25 & \textbf{0.80(0.02)} & 0.81(0.02) & 0.81(0.02) & 0.81(0.02) \\ 
   None & None & 50 & \textbf{0.80(0.01)} & \textbf{0.80(0.01)} & \textbf{0.80(0.01)} & \textbf{0.80(0.01)} \\ 
    &  & 100 & \textbf{0.80(0.01)} & \textbf{0.80(0.01)} & \textbf{0.80(0.01)} & \textbf{0.80(0.01)} \\ 
    \hline
    &  & 25 & 0.90(0.02) & \textbf{0.81(0.02)} & 0.93(0.02) & \textbf{0.81(0.02)} \\ 
   None & Spectral & 50 & 0.91(0.01) & \textbf{0.80(0.01)} & 0.92(0.01) & \textbf{0.80(0.01)} \\ 
    &  & 100 & 0.91(0.01) & \textbf{0.80(0.01)} & 0.93(0.01) & \textbf{0.80(0.01)} \\ 
  \hline
    &  & 25 & 1.03(0.02) & 0.92(0.02) & 1.00(0.03) & \textbf{0.87(0.04)} \\ 
   Translation & Spectral & 50 & 1.01(0.01) & 0.89(0.01) & 0.95(0.02) & \textbf{0.82(0.03)} \\ 
    &  & 100 & 1.02(0.01) & 0.89(0.01) & 0.97(0.01) & \textbf{0.83(0.04)} \\ 
  \hline
    &  & 25 & 0.98(0.03) & 0.91(0.02) & 0.96(0.03) & \textbf{0.86(0.03)} \\ 
   Diffeomorphism & Spectral & 50 & 0.97(0.01) & 0.88(0.01) & 0.93(0.01) & \textbf{0.83(0.02)} \\ 
    &  & 100 & 0.99(0.01) & 0.89(0.01) & 0.96(0.01) & \textbf{0.84(0.01)} \\ 
   \hline
    \end{tabular}
    \caption{MAD (standard error) estimates (in $\mu g/m^3$) for the proposed model with both smoothing and warping components, only smoothing component and only warping component along with a SLR model for different scenarios. The lowest MAD value in each case is in bold.}
    \label{tab:app_1}
\end{table}

\begin{table}[!h]
\setlength{\tabcolsep}{0 pt}
    \centering
    \begin{tabular}{>{\small}c@{\hspace{0 pt}} >{\small}c@{\hspace{1 pt}} >{$}c<{$}@{\hspace{7 pt}} c@{\hspace{7 pt}} c@{\hspace{7 pt}} c@{\hspace{7 pt}} c}
    \hline
         Warp&Smoothing&n& SLR& \multicolumn{3}{c}{Proposed Model}\\
         \cline{5-7}
         & & & & Warp &Smooth& Both\\
         \hline
  &  & 25 & 0.95(0.01) & 0.95(0.01) & 0.95(0.01) & 0.95(0.01) \\ 
   None & None & 50 & 0.95(0.00) & 0.95(0.00) & 0.95(0.00) & 0.95(0.00) \\ 
    &  & 100 & 0.95(0.00) & 0.95(0.00) & 0.95(0.00) & 0.95(0.00) \\ 
  \hline
    &  & 25 & 0.95(0.01) & 0.95(0.01) & 0.95(0.01) & 0.95(0.01) \\ 
   None & Spectral & 50 & 0.95(0.01) & 0.95(0.00) & 0.94(0.01) & 0.95(0.00) \\ 
    &  & 100 & 0.95(0.00) & 0.95(0.00) & 0.94(0.00) & 0.95(0.00) \\ 
  \hline
    &  & 25 & 0.95(0.01) & 0.95(0.01) & 0.94(0.01) & 0.95(0.01) \\ 
   Translation & Spectral & 50 & 0.95(0.00) & 0.95(0.00) & 0.95(0.01) & 0.95(0.01) \\ 
    &  & 100 & 0.95(0.00) & 0.95(0.00) & 0.95(0.00) & 0.95(0.00) \\ 
  \hline 
    &  & 25 & 0.95(0.01) & 0.94(0.01) & 0.94(0.01) & 0.95(0.01) \\ 
   Diffeomorphism & Spectral & 50 & 0.95(0.00) & 0.94(0.00) & 0.95(0.00) & 0.95(0.00) \\ 
    &  & 100 & 0.95(0.00) & 0.94(0.00) & 0.95(0.01) & 0.95(0.00) \\ 
   \hline
    \end{tabular}
    \caption{Coverage (standard error) estimates for the proposed model with  only smoothing component, only warping component and the full model along with an SLR model for different scenarios.}
    \label{tab:app_2}
\end{table}

\subsection{Additional Figures for Day by Day Comparison of Models in the Data Analysis}\label{subsec: dat_fig_extra}

We present additional images from data analysis here. The Figure \ref{fig:app_mad_mov} shows the performance of the four models, as in Section \ref{s:data}, for every run (each run being based on each day) based on the metrics MAD and coverage. The inference is similar to that in Section \ref{s:data}.

\begin{figure}[!h]
    \begin{subfigure}{0.49\textwidth}
    \centering
    \includegraphics[height = 0.3\textheight]{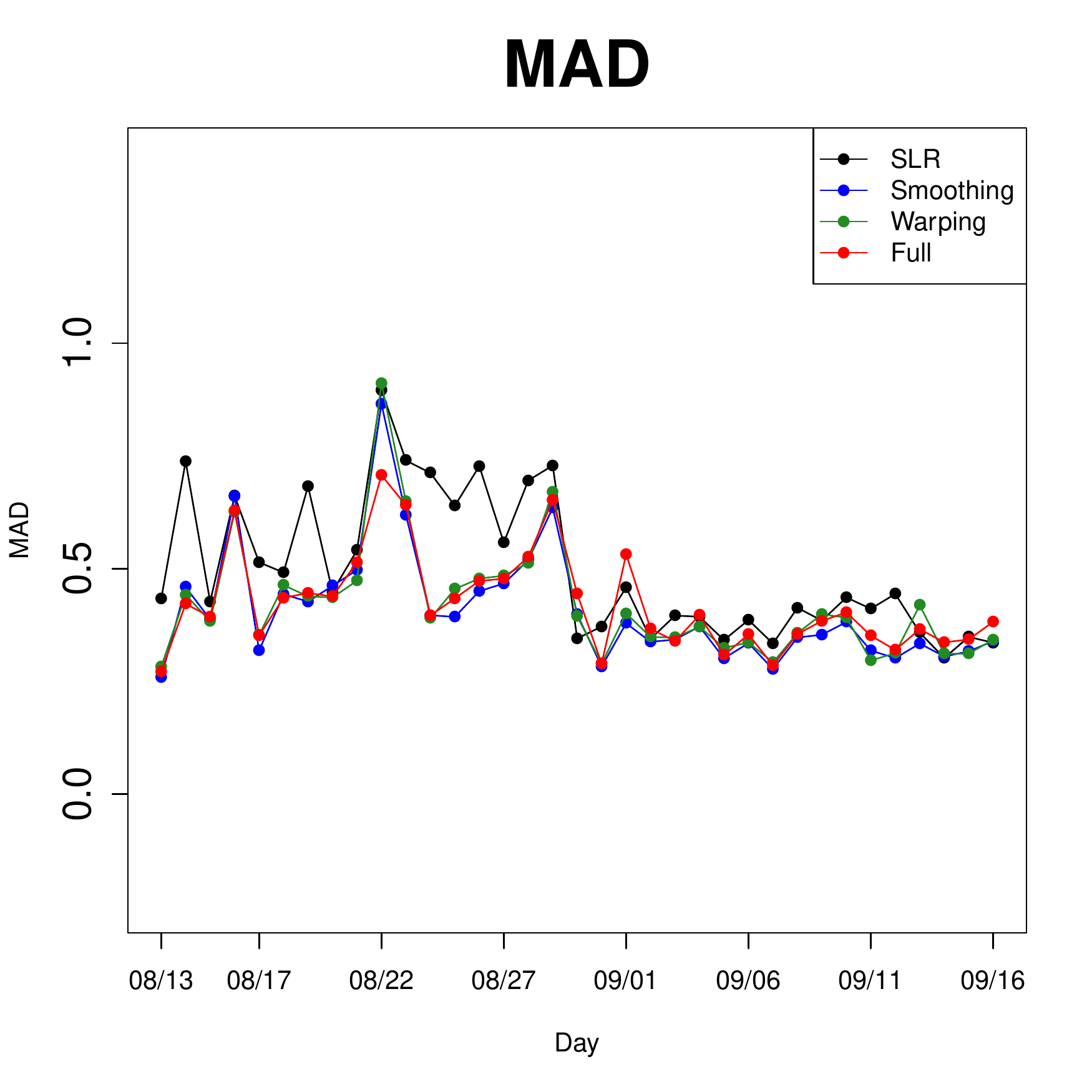}
\end{subfigure}
\begin{subfigure}{0.49\textwidth}
    \centering
    \includegraphics[height = 0.3\textheight]{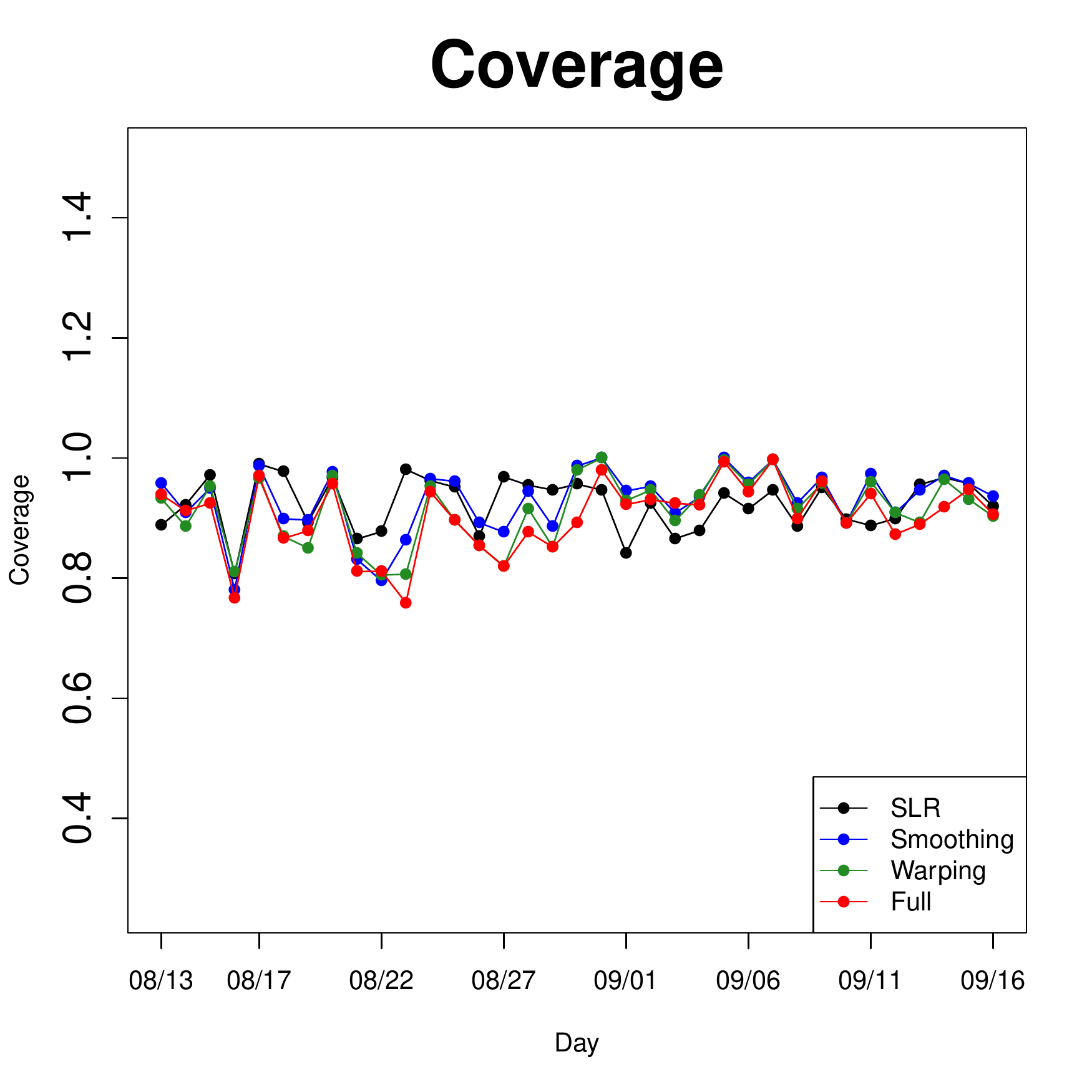}
\end{subfigure}
    \caption{Daily prediction MAD (left panel) in $\mu g/m^3$ and coverage (right panel) for the four models}
    \label{fig:app_mad_cov}
\end{figure}

\bibliographystyle{imsart-nameyear}
\bibliography{warping}

\end{document}